\documentclass[preprint,12pt,authoryear]{elsarticle}

\usepackage{hyperref}
\usepackage{graphicx}
\usepackage{amssymb}
\usepackage{amsmath}
\usepackage{amsthm}
 \usepackage{natbib}
\usepackage{orcidlink}
\usepackage{booktabs}
\usepackage{subfigure}
\usepackage{cancel}
\usepackage{graphicx}
\usepackage{multirow} 
\usepackage{threeparttable}
\usepackage{longtable}
\usepackage{multicol}
\usepackage{longtable}
\usepackage{array}
\usepackage{booktabs}
\usepackage{caption}

\journal{Journal of High Energy Astrophysics}

\begin{document}

\begin{frontmatter}

\title{Hurst index of gamma-ray burst light curves and its statistical study}

\author[label1]{Ruo-Yu Guan\orcidlink{0009-0000-8680-1762}}
\affiliation[label1]{organization={Department of Astronomy, School of Physics, Huazhong University of Science and Technology},
            city={Wuhan},
            postcode={430074}, 
            country={China}}

\author[label2]{Feifei Wang} 
\affiliation[label2]{organization={School of Mathematics and Physics, Qingdao University of Science and Technology},
	city={Qingdao},
	postcode={266061}, 
	country={China}}

\author[label1]{Yuan-Chuan Zou\orcidlink{0000-0002-5400-3261}\corref{cor1}}
\cortext[cor1]{Corresponding author: Yuan-Chuan Zou, zouyc@hust.edu.cn}

\begin{abstract}
Gamma-ray bursts (GRBs) rank among the most powerful astrophysical phenomena, characterized by complex and highly variable prompt emission light curves that reflect the dynamics of their central engines. In this work, we analyze a sample of 163 long-duration GRBs detected by the Burst and Transient Source Experiment (BATSE), applying detrended fluctuation analysis (DFA) to derive the Hurst index as a quantitative descriptor of temporal correlations in the light curves. We further explore statistical correlations between the Hurst index and 12 other observational parameters through regression and correlation analyses. Our results reveal anti-correlations between the Hurst index and the burst durations ($T_{50}$, $T_{90}$), and moderate positive correlations with peak photon flux proxies (P$_{pk1}$--P$_{pk3}$). By contrast, the standard spectral parameters (including the low-energy index $\alpha$) show no evidence for a linear dependence on the Hurst index in our sample. We do not find a clear monotonic weakening of the correlation strength from 64\,ms to 1024\,ms peak-flux measures; rather, the correlation coefficients for P$_{pk1}$--P$_{pk3}$ are comparable within uncertainties. The results offer new perspectives on the temporal structure of the GRB emission and its potential link to the underlying physical mechanisms driving these bursts.

\end{abstract}


\end{frontmatter}

\section{Introduction} \label{sec:intro}

Gamma-ray bursts (GRBs) are the most powerful and luminous transient events in the universe, believed to originate from relativistic jets produced by catastrophic astrophysical processes. Observationally, GRBs are broadly classified into two categories---long and short---based on a characteristic duration threshold of approximately two seconds\citep{1993ApJ...413L.101K}. Long GRBs ($T_{90}>2\,\mathrm{s}$) are generally associated with the core collapse of massive, rapidly rotating stars\citep{1993ApJ...405..273W, 1998ApJ...494L..45P} and are frequently accompanied by broad-lined Type Ic supernovae\citep{2006ARA&A..44..507W}. In contrast, short GRBs ($T_{90}<2\,\mathrm{s}$) are commonly linked to the mergers of compact binary systems\citep{1998ApJ...507L..59L}, such as neutron star--neutron star\citep{2017ApJ...848L..13A, 2017ApJ...848L..14G, 1986ApJ...308L..43P} or neutron star--black hole systems\citep{1992ApJ...397..570M, 1991AcA....41..257P}.

The light curves (LCs) of the GRBs serve as a vital diagnostic of the underlying physical processes governing these extreme astrophysical explosions and the conditions in the vicinity of their central engines. Characterized by highly variable, non-thermal emission that spans timescales from milliseconds to several minutes \citep{2012MNRAS.425L..32M}, GRB light curves display a remarkable range of temporal structures, including single-peaked components, multi-episodic pulses, and intricate patterns of variability across different energy bands. This morphological diversity indicates the complex nature of relativistic outflows and energy dissipation mechanisms, which may involve internal shocks, magnetic reconnection, or cascades driven by turbulence within the jet \citep{2011ApJ...726...90Z}. 

Over the years, numerous indicators have been developed to quantify the temporal variability of GRBs. These include pulse decomposition analysis \citep{Norris1996}, variability indices \citep{Fenimore2000}, Fourier power density spectra \citep{Beloborodov2000}, autocorrelation functions \citep{Borgonovo2004}, and minimum variability time scales \citep{2012MNRAS.425L..32M}. Crucially, some studies have found significant correlations between variability indicators and other properties such as luminosity \citep{Reichart2001,Guidorzi2005}, 
suggesting that the light curve structure may provide key diagnostics of jet composition and energy dissipation.

Originally introduced by \citet{1994PhRvE..49.1685P, 1995Chaos...5...82P}, detrended fluctuation analysis (DFA) is a fundamental method for examining the scalar properties in a variety of time series datasets. Later, \citet{2002PhyA..316...87K} expanded DFA to analyze multifractal processes, resulting in the multifractal detrended fluctuation analysis (MFDFA) technique. A generalized Hurst index was defined to quantify the long-range dependence of a time series, initially introduced by \citet{Hurst1951} in the study of the long-term storage capacity of reservoirs. This method effectively addresses correlation issues in time-series data and is applicable to both discrete and continuous stochastic processes.
It has been used successfully in a wide range of applications spanning several fields, such as music \citep{2007JSMTE..04...12J}, heartbeat dynamics \citep{1999Natur.399..461I}, electroencephalogram sleep data \citep{2020Chaos..30g3138P, 2021EPJP..136...10P}, arterial pressure \citep{2020CNSNS..8505232P}, cosmic microwave radiation \citep{2011PhRvE..84b1103M, 2013MNRAS.434.3597M}, atmospheric turbulence effects on stellar images \citep{2014OptL...39.3718Z}, sunspot fluctuations \citep{2006JSMTE..02..003S, 2009JSMTE..02..066H}, solar flares \citep{2020SoPh..295..123L}, gravitational wave detection \citep{2018ApJ...864..162E}, quasiperiodic oscillation searching \citep{2021ApJ...911...20T}, blazars \citep{2020ApJS..250....1T}, and fast radio bursts \citep{2023ApJ...949L..33W}.
In this context, DFA and its extension to MFDFA offer powerful tools to characterize temporal correlations and long-range memory in GRB light curves. Unlike traditional variability measures, DFA quantifies the scaling behavior and persistence of non‑stationary time series, making it particularly suited to GRB data \citep{1994PhRvE..49.1685P,2002PhyA..316...87K,2014OptL...39.3718Z}.

Since its launch in 1991 onboard NASA's Compton Gamma Ray Observatory, the Burst and Transient Source Experiment (BATSE) has been instrumental in shaping our current understanding of GRBs \citep{1993ApJ...413..281B}. One of BATSE's most impactful achievements is its systematic collection of high-resolution light curves in four well-defined energy bands: 20-50 keV, 50-100 keV, 100-300 keV, and above 300 keV, which has facilitated detailed multiband temporal and spectral investigations of GRB prompt emission. Using this extensive data set, \citet{2021ApJ...919...37H} found that most GRB pulses can be well characterized by a smooth, single-peaked emission profile accompanied by a temporally symmetric residual component, shedding new light on the structure and physical origin of GRB emission mechanisms. We also chose the BATSE data for our analysis because of its high sensitivity.

In this paper, we analyze the detrended fluctuations in the light curves of 163 long GRBs detected by BATSE, derive a characteristic index (the Hurst index) from each GRB light curve treated as a time series, and perform regression and correlation analyses between this index and 12 different physical parameters that have been observed for these long GRBs. The paper is arranged as follows. In Section \ref{sec:data_analysis}, we present our data sample and describe the 12 physical parameters for which we performed the correlation analysis. The detrended fluctuation analysis model, along with its parameter selection and statistical analysis, is presented in Section \ref{sec:method}. The results and conclusions are reported in Sections \ref{sec:results} and \ref{sec:discussion_conclusions}, respectively.

\section{Data Analysis} \label{sec:data_analysis}
\subsection{Sample Selection} \label{subsec:sample}

Our analysis is based on a carefully selected sample of 163 long-duration GRBs detected by BATSE, corresponding to trigger numbers ranging from 107 to 1997\footnote{\href{https://heasarc.gsfc.nasa.gov/FTP/compton/data/batse/ascii_data/64ms/}{https://heasarc.gsfc.nasa.gov/FTP/compton/data/batse/ascii\_data/64ms/}}, which is the intersection of the BATSE Gamma Ray Burst Lightcurve Image Archive\footnote{\href{https://gammaray.nsstc.nasa.gov/batse/grb/lightcurve/}{https://gammaray.nsstc.nasa.gov/batse/grb/lightcurve/}} and the CGRO/BATSE 4B Catalog\footnote{\href{https://heasarc.gsfc.nasa.gov/W3Browse/cgro/batse4b.html}{https://heasarc.gsfc.nasa.gov/W3Browse/cgro/batse4b.html}}. The sample selection criteria also included: complete determinations of the spectral parameters from band function fittings and availability of all 12 other physical parameters under investigation from \citet{Wang2020}. The 12 selected parameters are shown in Table \ref{tab:para}. The DISCSC light curves are originally provided at 64~ms resolution. In this work, we rebin the DISCSC data to a uniform time resolution of 1024~ms by summing 16 consecutive 64~ms bins, so that all bursts are analyzed on the same temporal grid. This uniform binning is adopted because DFA results are sensitive to the sampling resolution; fixing the bin width across the sample avoids introducing an additional burst-dependent scale that could bias the distribution of Hurst indices and compromise the comparability required for our statistical study.

This choice also imposes a practical requirement on the minimum number of data points available for each burst. Starting from an initially larger candidate set, we excluded 36 GRBs with $T_{90}<8.6$~s, for which the rebinned light curves would contain too few 1024~ms bins within the prompt-emission window to yield a meaningful DFA scaling fit. After this filtering, 163 long GRBs remain and constitute our final sample. We note that relatively short bursts (e.g., those with $T_{90}\lesssim 30$~s) are still included as long as they satisfy the above minimum-length requirement under the uniform 1024~ms binning.

\begin{figure}[htbp]
    \centering
    \includegraphics[width=1\linewidth]{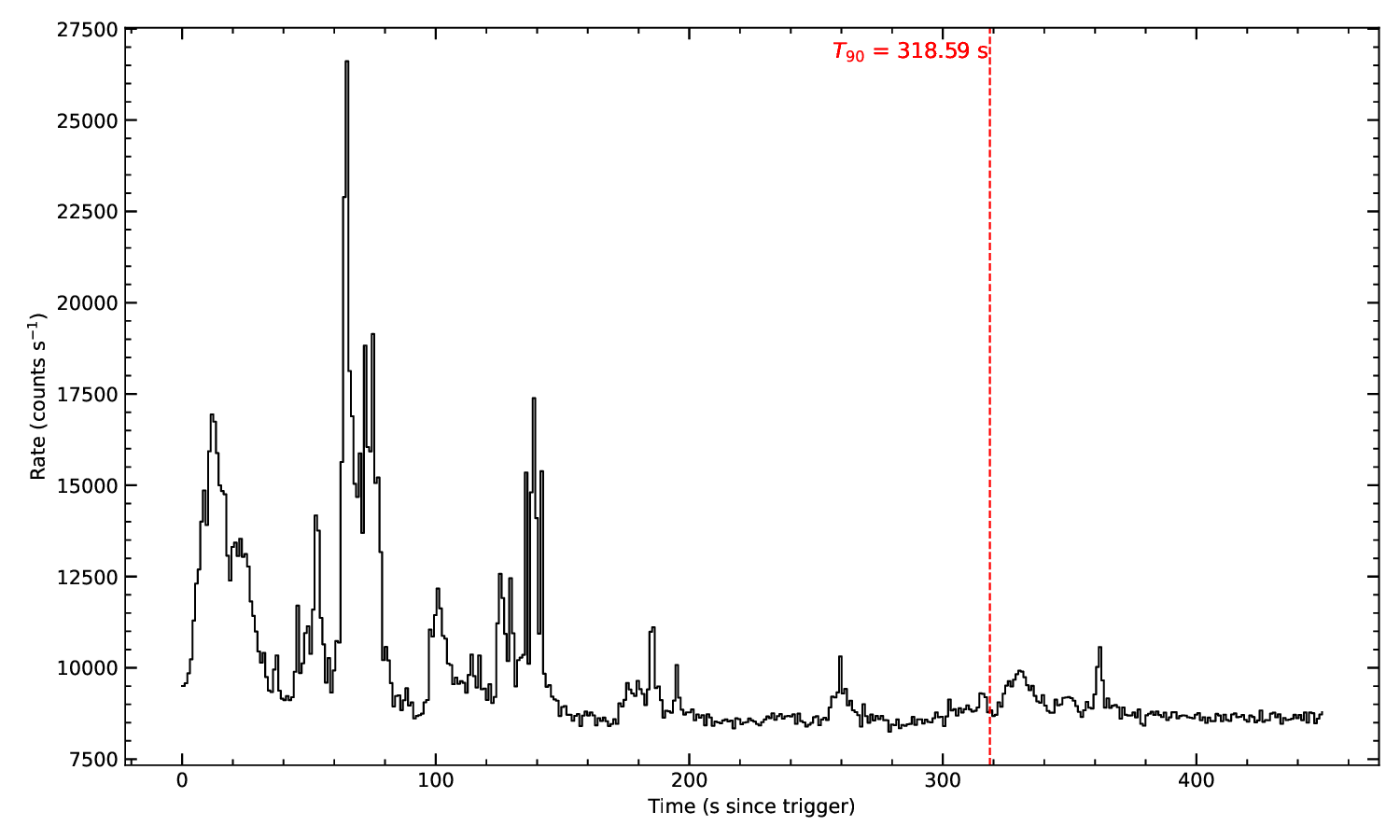}
    \caption{Light curve of GRB 920110A as an example, the energy channel is 1-4 ($>20$ keV).} 
    \label{fig:LC}
\end{figure}

The time interval for each burst is selected from the trigger time up to $T_{90}$. This standardization provides a uniform and reproducible analysis window, guaranteeing that the scaling properties derived from detrended fluctuation analysis are representative of the burst's intrinsic variability during its most active phase and enabling a meaningful comparison of Hurst indices for robust statistical evaluation. Fig. \ref{fig:LC} presents the light curve of GRB~920110A as an example, one of the events included in our sample, over the time span corresponding to its \( T_{90} \) duration. \ref{appendix} (\ref{tab:list_HI}) lists the Hurst indices for all GRBs in our sample; the corresponding physical parameters used in the correlation analysis are taken from \citet{Wang2020}. We emphasize that in this work $T_{90}$ is used as a practical duration parameter to define an analysis window that covers the vast majority of the prompt-emission variability, rather than as a strict physical gate that begins at the canonical $t_{\rm start}$ of the $T_{90}$ definition. In many BATSE long GRBs, the light curve shows a clear rise and non-negligible variability immediately after trigger ($t\approx 0$), and excluding the early segment prior to the onset of the cataloged $T_{90}$ interval would remove part of the burst structure that can contribute to the scaling behavior measured by DFA. Therefore, we adopt the interval $t\in[0,\,T_{90}]$ for all bursts to ensure methodological uniformity across the sample and to avoid selectively discarding early-time variability. The purpose of our study is to obtain a standard set of DFA-based scaling measures for statistical comparison across the sample. In Fig.~\ref{fig:LC}, the dashed line marks the endpoint of the adopted analysis window at $t=T_{90}$ (measured from the trigger). The caption explicitly clarifies that $T_{90}$ is a duration parameter, and that the vertical marker indicates the end of the $[0,\,T_{90}]$ window rather than the $[T_{5},\,T_{95}]$ window. For five triggers (No. 107, 110, 114, 211, 351) whose publicly available DISCSC records terminate before $t=T_{90}$, we perform DFA on the actually available window $[0,\,t_{\rm end}^{\rm data}]$ and treat the cataloged $T_{90}$ as a reference duration parameter.

\subsection{Physical Parameters} \label{subsec:phys_para}

We analyze 12 key physical parameters that characterize different aspects of GRB emission. These parameters can be broadly categorized into three groups: (1) spectral parameters describing the energy distribution of photons, (2) temporal parameters characterizing the burst duration and variability, and (3) flux-related parameters measuring the burst intensity. Table \ref{tab:para} summarizes these parameters with their units and physical interpretations. The spectral parameters $\alpha$ and $\beta$ represent the low- and high-energy photon indices from the Band function fits, while $E_{peak}$ corresponds to the peak energy in the $\nu F_\nu$ spectrum. The temporal parameters $T_{50}$ and $T_{90}$ describe the durations of the bursts that contain total counts of 50\% and 90\%, respectively, and V quantifies the degree of fluctuations in the light curve. The flux-related parameters include fluence ($F_g$), peak flux ($F_{pk}$), and photon flux ($P_{pk}$) measured in different time bins. The hardness ratio (HR) provides information about the spectral hardness of the burst.
All data were taken from \citet{Wang2020}, while they were gathered from the references therein. 

\begin{table}
    \centering
    \resizebox{\linewidth}{!}{
    \begin{tabular}{ccc}
    \hline
    \hline
    Parameter & Units & Description  \\
    \hline
    $\alpha$     & -- & Opposite value of the low-energy spectrum index of the Band model \\
    $\beta$     & -- & Opposite value of the high-energy spectrum index of the Band model \\
    $E_{peak}$     & keV &Spectral peak energy of the Band model \\
    $F_g$     & $10^{-6}$ erg cm$^{-2}$ & Fluence in the 20-2000 keV energy band \\
    $F_{pk}$     & $10^{-6}$ erg cm$^{-2}$ s$^{-1}$ & Peak flux with 1 s time bin in the observer's frame 1-10000 keV energy band \\
    HR     & -- & Hardness ratio between 100-2000 keV and 20-100 keV \\
    $P_{pk1}$     & ph cm$^{-2}$ s$^{-1}$ & Peak photon flux in the 64 ms time bin of 10-1000 keV \\
    $P_{pk2}$     & ph cm$^{-2}$ s$^{-1}$ & Peak photon flux in the 256 ms time bin of 10-1000 keV \\
    $P_{pk3}$     & ph cm$^{-2}$ s$^{-1}$ & Peak photon flux in the 1024 ms time bin of 10-1000 keV \\
$T_{50}$     & s & Burst duration between 25\% and 75\% of the cumulative counts (following \citet{Wang2020}) \\
$T_{90}$     & s & Burst duration between 5\% and 95\% of the cumulative counts (following \citet{Wang2020}) \\
    V (variability)     & -- & Light curve variability index based on \citet{2000astro.ph..4176F}, \\ & & which also corresponds to variability1 in \citet{Wang2020} \\
    \hline
    \end{tabular}
    }
    \caption{Description of the 12 physical parameters analyzed in this study. The parameters are grouped into spectral ($\alpha$, $\beta$, $E_{peak}$, HR), temporal ($T_{50}$, $T_{90}$, V), and flux-related ($F_g$, $F_{pk}$, $P_{pk1-3}$) categories. More detailed description can be found in \citet{Wang2020}.}
    \label{tab:para}
\end{table}

\section{Methods}\label{sec:method}

\subsection{Detrended Fluctuation Analysis (DFA)}\label{subsec:DFA}

Detrended fluctuation analysis (DFA) was originally proposed by \citet{1994PhRvE..49.1685P} and later generalized to multifractal detrended fluctuation analysis (MF-DFA) by \citet{2002PhyA..316...87K}. In this work, we focus on the monofractal case (DFA) by fixing $q=2$ in the MF-DFA formalism.

Given a GRB light curve represented by a discrete time series $x(i)$ of length $N$, DFA proceeds as follows:

(1) \textit{Profile construction.} We construct the cumulative (integrated) profile
\begin{equation}
    y(l)=\sum_{i=1}^{l}\left[x(i)-\langle x\rangle\right],
\end{equation}
where $\langle x\rangle$ is the mean of the time series. Here $x(i)$ is the binned photon counts (counts per bin) in the $i$-th time bin of width $\Delta t = 1.024~\mathrm{s}$ (after summing the four DISCSC channels).

(2) \textit{Segmentation and detrending.} For a chosen scale (window size) $s$, the profile $y(l)$ is divided into $N_s=\lfloor N/s\rfloor$ non-overlapping segments of length $s$. To reduce boundary effects and to use the entire time series, we perform the same segmentation starting from the end of the profile, yielding a total of $2N_s$ segments \citep{2002PhyA..316...87K}. In each segment $\nu$, we fit an $n$-th order polynomial trend $y_\nu^{(n)}(l)$ (in this work $n=1$) via least squares and subtract it from the profile.

(3) \textit{Local variance.} For each segment $\nu$, the detrended variance is computed as
\begin{equation}
    F^2(\nu,s)=\frac{1}{s}\sum_{k=1}^{s}\left[y_\nu(l)-y_\nu^{(n)}(l)\right]^2,
\end{equation}
where $y_\nu(l)$ denotes the profile values restricted to segment $\nu$.

(4) \textit{Fluctuation function.} The fluctuation function at scale $s$ is obtained by averaging over all $2N_s$ segments:
\begin{equation}
    F(s)=\left[\frac{1}{2N_s}\sum_{\nu=1}^{2N_s}F^2(\nu,s)\right]^{1/2}.
\end{equation}

(5) \textit{Scaling and slope estimation.} Repeating the above steps for a set of scales $\{s\}$, DFA predicts a power-law scaling
\begin{equation}
    F(s) \propto s^{k},
\end{equation}
where $k$ is the fitted slope in the $\ln F(s)$--$\ln s$ plane. We estimate $k$ using ordinary least squares (OLS) regression over all available scales used in the computation (see Section~\ref{subsec:implementation}). To avoid confusion with the low-energy spectral index $\alpha$ of the Band function, we denote the DFA scaling slope by $k$ throughout this paper.

Following the convention implemented in the \texttt{MFDFA} package \citep{2022CoPhC.27308254R}, we report the Hurst index for the light-curve series as
\begin{equation}\label{eq:H}
    H = k-1.
\end{equation}
Here $H$ is an empirical DFA scaling index defined by Eq. (\ref{eq:H}). For finite-length and strongly nonstationary burst light curves, the fitted scaling exponent may fall outside the canonical range expected for idealized fGn/fBm, and such cases should be interpreted with caution.

\subsection{Implementation and analysis configuration}\label{subsec:implementation}

\paragraph{Time resolution and rebinning}
The original BATSE light-curve data used in this work have a native time bin width of 64~ms.
Prior to the DFA computation, we rebin each burst to a uniform time resolution of 1024~ms (1.024~s) by summing consecutive 16 bins.
For a discrete count series $x_{64}(i)$ sampled at 64~ms, the rebinned series is
\begin{equation}
x_{1024}(j)=\sum_{m=1}^{16} x_{64}\left(16(j-1)+m\right)\quad j=1, 2, \dots, N_{1024}.
\end{equation}
All Hurst indices reported in this paper are derived from the rebinned 1024~ms light curves; the 64~ms data are used only as the input for rebinning. We note that BATSE monitors triggering on 64~ms, 256~ms, and 1024~ms timescales; this triggering timescale does not imply that our DFA is performed on 64~ms bins.

\paragraph{Time window definition}
For each GRB, we extract the DISCSC time series from the trigger epoch ($t=0$) to $t=T_{90}$, where $T_{90}$ is the catalog duration parameter. We emphasize that, in the BATSE definition, the physical $T_{90}$ interval corresponds to the time span between the 5\% and 95\% accumulation levels of the background-subtracted cumulative counts (i.e., $T_{5}$ to $T_{95}$) \citep{1995ApJ...439..542N}. In our implementation, we use $[0, T_{90}]$ as a uniform and reproducible analysis window referenced to the trigger time; we refer to this as the ``0--$T_{90}$ window'' to avoid confusion with the catalog $T_{5}$--$T_{95}$ interval.

\paragraph{Scale selection and detrending order}
We use logarithmically spaced scales from $s_{\min}=3$ bins to $s_{\max}=\lfloor N/10\rfloor$ bins, with 200 candidate scale points before integer rounding and deduplication. With $\Delta t=1.024$~s, the minimum scale corresponds to $s_{\min}\Delta t=3.072$~s. We adopt first-order polynomial detrending in each segment. For short rebinned series where the nominal upper scale $s_{\max}=\lfloor N/10\rfloor$ falls below $s_{\min}=3$ bins, we retain the burst by fitting over the available integer scales after rounding and deduplication.

\paragraph{Hurst index and uncertainty}
For each GRB light curve, we evaluate the fluctuation function $F(s)$ on a set of logarithmically spaced window sizes $s$ and determine the scaling exponent by fitting the linear model
\begin{equation}
\ln F(s) = k\,\ln s + b
\end{equation}
using an unweighted ordinary least-squares (OLS) regression over all adopted scales. We then report the Hurst index as $H = k - 1$.

When the regression is well defined, the statistical uncertainty of the fitted slope $k$ is quantified by its OLS standard error. Specifically, let $n$ be the number of scale points used in the fit, $y_i=\ln F(s_i)$, and $\hat{y}_i$ the fitted values. The residual sum of squares is
\begin{equation}
\mathrm{RSS}=\sum_{i=1}^{n}\left(y_i-\hat{y}_i\right)^2,
\end{equation}
and the mean squared error (i.e., the unbiased estimator of the residual variance) is
\begin{equation}
\mathrm{MSE}=\frac{\mathrm{RSS}}{n-2},
\end{equation}
where $n-2$ is the number of degrees of freedom for a linear fit (slope and intercept). With the design matrix $X=[\ln s_i,\,1]$, the standard error of the slope is
\begin{equation}
\mathrm{SE}(k)=\sqrt{\mathrm{MSE}\,\left[(X^{\mathsf T}X)^{-1}\right]_{11}}.
\end{equation}
Because $H$ differs from $k$ by a constant offset, the propagated uncertainty is simply $\mathrm{SE}(H)=\mathrm{SE}(k)$.

In a small number of cases (typically very short time series), the regression-based uncertainty may be ill-conditioned or numerically unstable. For these bursts, we estimate the uncertainty from repeated resampling of the time series (bootstrap or subsampling, depending on data length) and record the adopted uncertainty-estimation method for each GRB.

\begin{figure}
    \centering
    \includegraphics[width=0.7\linewidth]{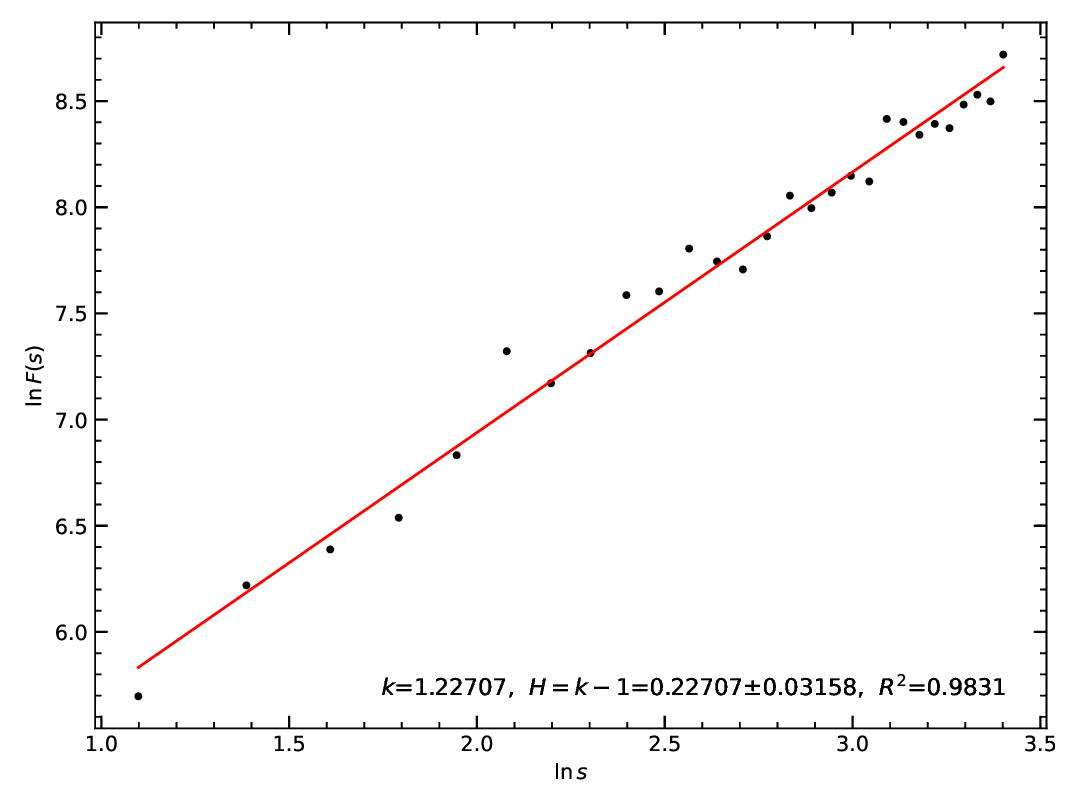}
    \caption{$\ln F(s)-\ln s$ Scaling Relation and the OLS fit for GRB 920110A}
    \label{fig:dfa_example}
\end{figure}

\paragraph{Worked example}
To make the derivation of $H$ explicit, Figure~\ref{fig:dfa_example} shows the $\ln F(s)$--$\ln s$ scaling relation and the OLS fit for the GRB in Figure~\ref{fig:LC} (GRB~920110A). The corresponding $(s,F(s))$ data used for the fit are exported from our pipeline to enable full reproducibility.

\section{Results} \label{sec:results}

Based on the Hurst indices ($H$) measured for 163 BATSE GRBs (Section~\ref{subsec:implementation}), we examine their statistical association with 12 fundamental GRB observables compiled by \citet{Wang2020}. Following the approach of \citet{Wang2020} and the classical definition of the Pearson correlation coefficient \citep{1895RSPS...58..240P}, we use the Pearson $r$ to quantify the strength of each $H$--$x$ relation. Beyond the point estimate of $r$, we (i) propagate measurement uncertainties to obtain an uncertainty interval for $r$, (ii) assess the adequacy of a linear description, and (iii) evaluate whether a linear trend is preferred over a null (no-dependence) model.

The uncertainty in $r$ is propagated from measurement errors via Monte Carlo (MC) sampling. For each observable $x$ and the Hurst index $H$, we generate $N_{\rm MC}$ synthetic realizations of the dataset by drawing, for each burst,
$x_i' \sim \mathcal{N}(x_i,\sigma_{x,i})$ and $H_i' \sim \mathcal{N}(H_i,\sigma_{H,i})$,
and compute the corresponding Pearson coefficient $r'$ for each realization. We then summarize the resulting $r'$ distribution by reporting its 16th and 84th percentiles as the central 68\% interval, which we adopt as the MC-based uncertainty range of $r$.

To test whether a linear relation provides an adequate description of the data, we perform a linear regression that accounts for uncertainties in both coordinates and quantifies the goodness of fit using the reduced chi-square, $\chi_\nu^2$. Finally, to compare a linear trend against a null model in which $H$ is independent of $x$, we compute the Akaike and Bayesian information criteria (AIC and BIC) for both models and report the differences
\begin{equation}
\Delta \mathrm{AIC} \equiv \mathrm{AIC}_{\rm null}-\mathrm{AIC}_{\rm lin}, \qquad
\Delta \mathrm{BIC} \equiv \mathrm{BIC}_{\rm null}-\mathrm{BIC}_{\rm lin},
\end{equation}
so that $\Delta \mathrm{AIC}>0$ and $\Delta \mathrm{BIC}>0$ indicate a preference for the linear model over the null model after accounting for model complexity. The results are presented below.

\begin{figure}
    \centering
    \includegraphics[width=0.7\linewidth]{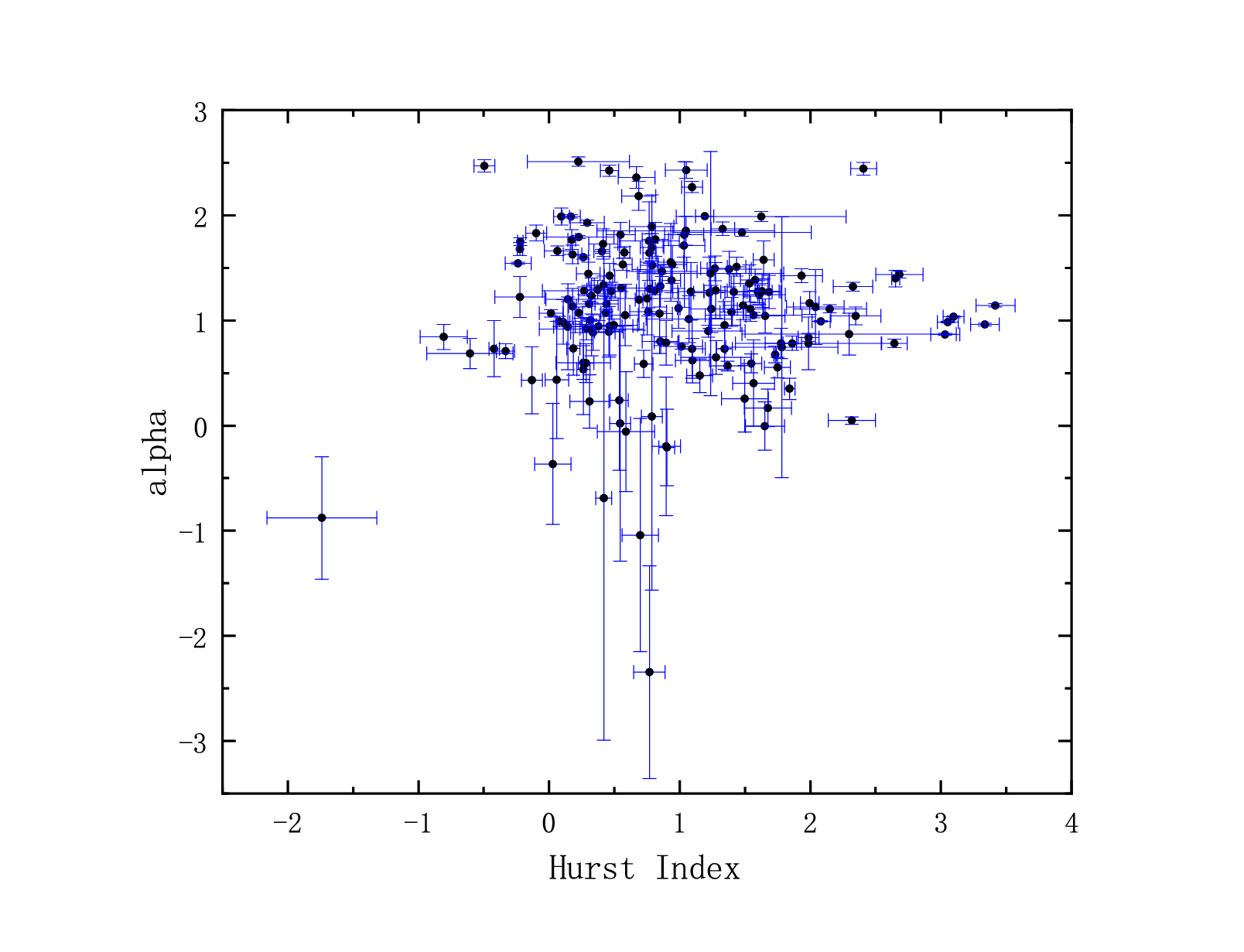}
    \caption{Scatter plot for $\alpha$ and Hurst index. For this parameter, the AIC/BIC model comparison favors the null (constant) model over the linear model; therefore, no regression line is shown.} The description of each parameter is in Section \ref{subsec:phys_para}.
    \label{fig:alpha}
\end{figure}

\begin{figure}
    \centering
    \includegraphics[width=0.7\linewidth]{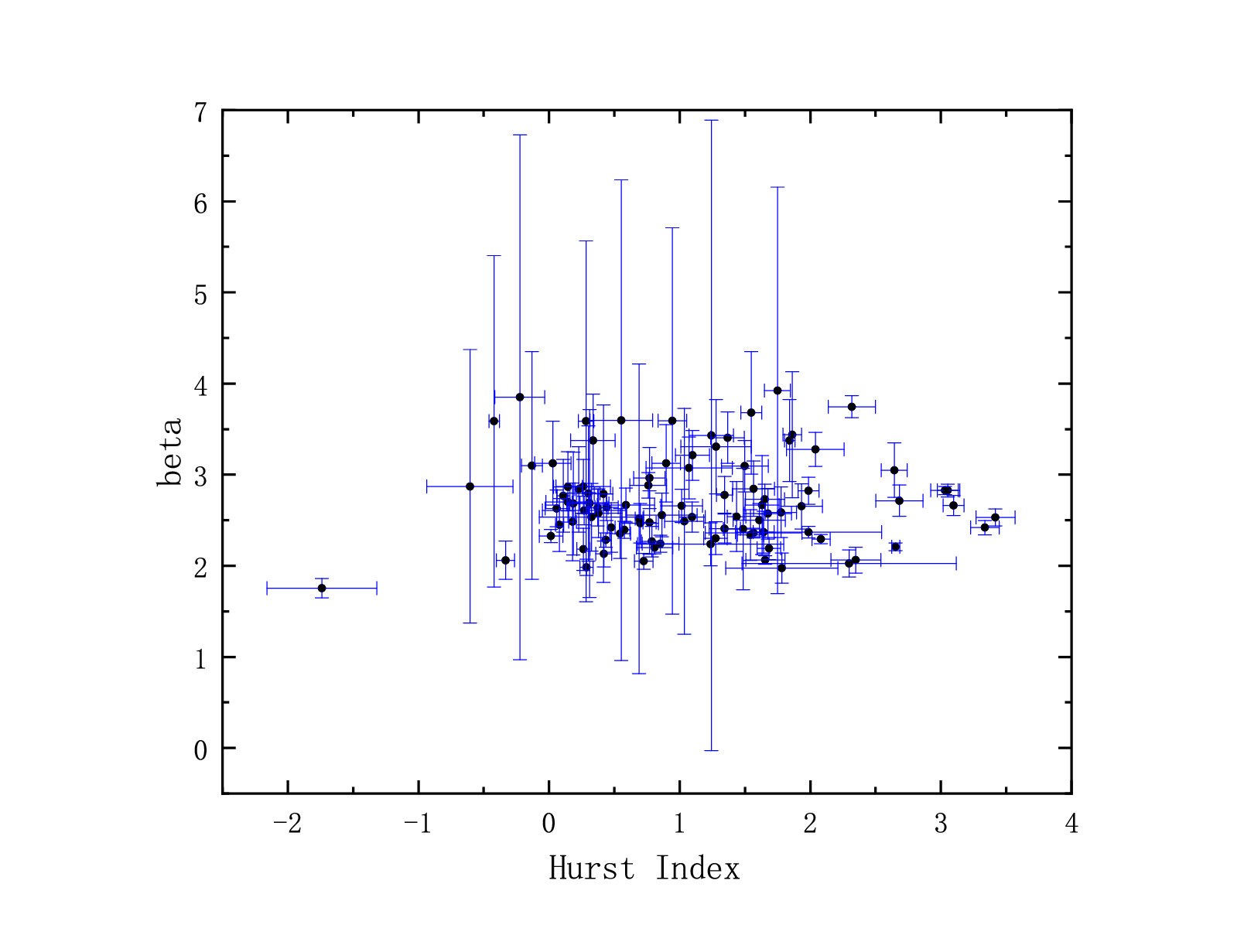}
    \caption{Scatter plot for $\beta$ and Hurst index. For this parameter, the AIC/BIC model comparison favors the null (constant) model over the linear model; therefore, no regression line is shown.} The description of each parameter is in Section \ref{subsec:phys_para}.
    \label{fig:beta}
\end{figure}

\begin{figure}
    \centering
    \includegraphics[width=0.7\linewidth]{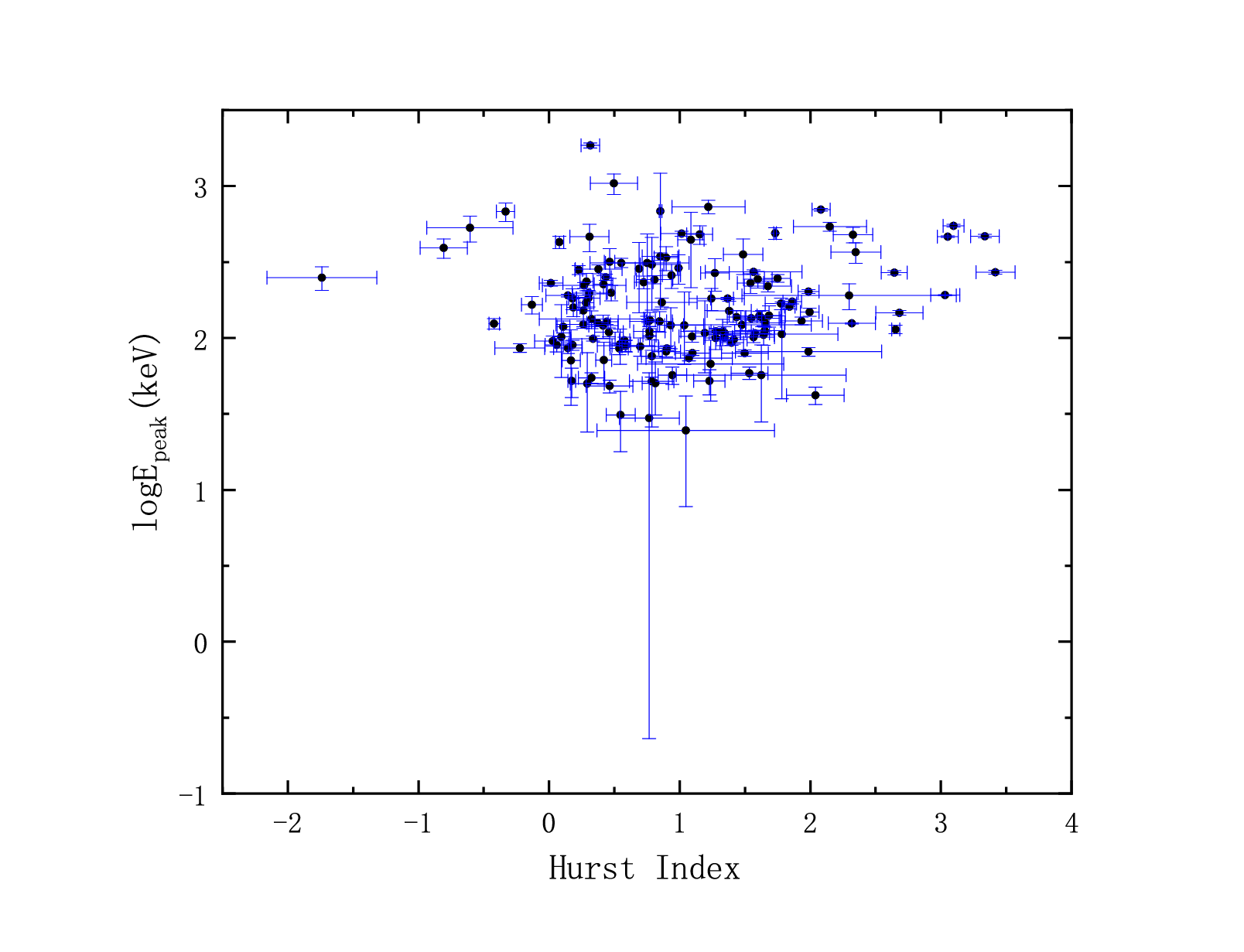}
    \caption{Scatter plot for $\log{E_{peak}}$ and Hurst index. For this parameter, the AIC/BIC model comparison favors the null (constant) model over the linear model; therefore, no regression line is shown.} The description of each parameter is in Section \ref{subsec:phys_para}.
    \label{fig:E_peak}
\end{figure}

\begin{figure}
    \centering
    \includegraphics[width=0.7\linewidth]{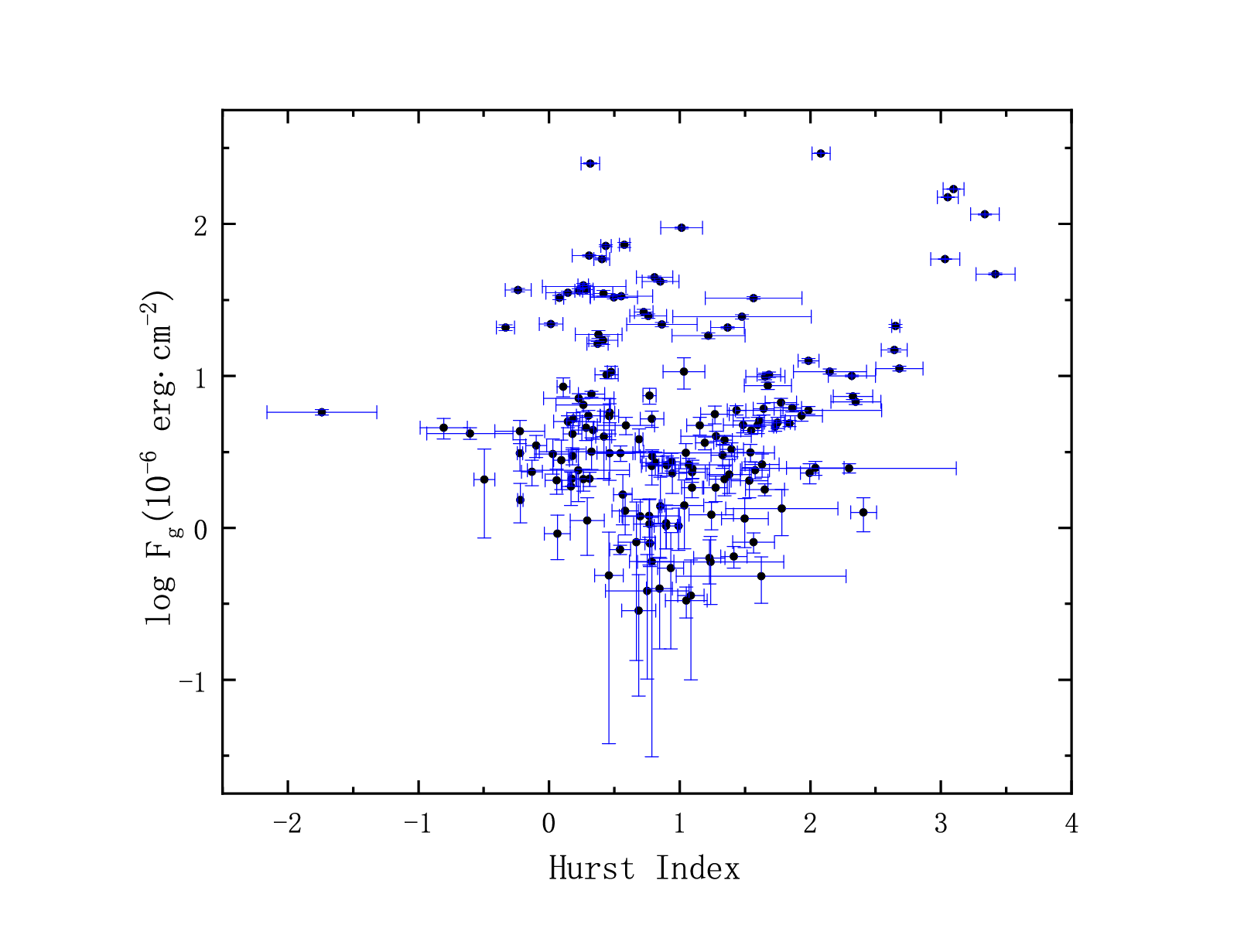}
    \caption{Scatter plot for $\log{F_g}$ and Hurst index. For this parameter, the AIC/BIC model comparison favors the null (constant) model over the linear model; therefore, no regression line is shown.} The description of each parameter is in Section \ref{subsec:phys_para}.
    \label{fig:F_g}
\end{figure}

\begin{figure}
    \centering
    \includegraphics[width=0.7\linewidth]{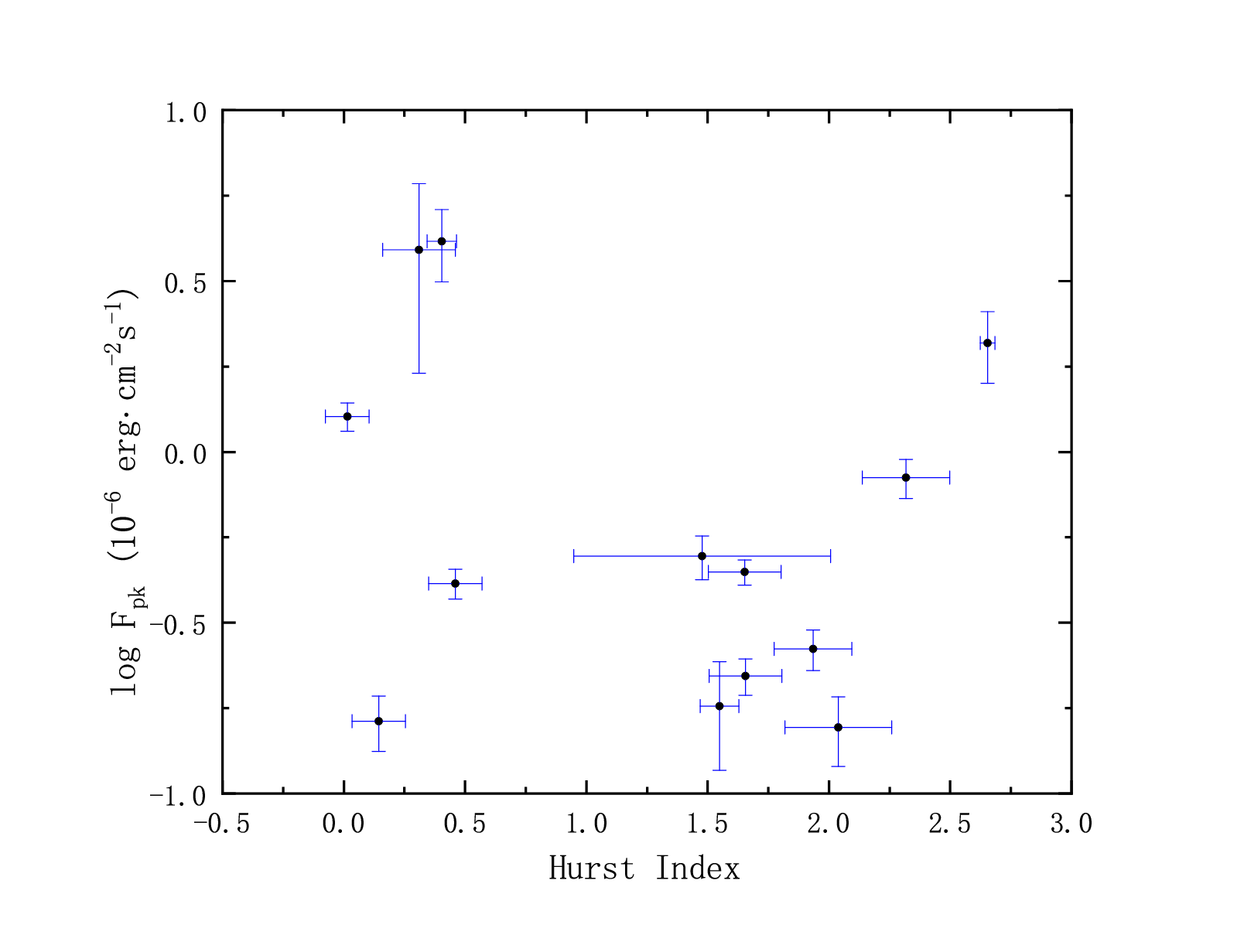}
    \caption{Scatter plot for $\log{F_{pk}}$ and Hurst index. For this parameter, the AIC/BIC model comparison favors the null (constant) model over the linear model; therefore, no regression line is shown.} The description of each parameter is in Section \ref{subsec:phys_para}.
    \label{fig:F_pk}
\end{figure}

\begin{figure}
    \centering
    \includegraphics[width=0.7\linewidth]{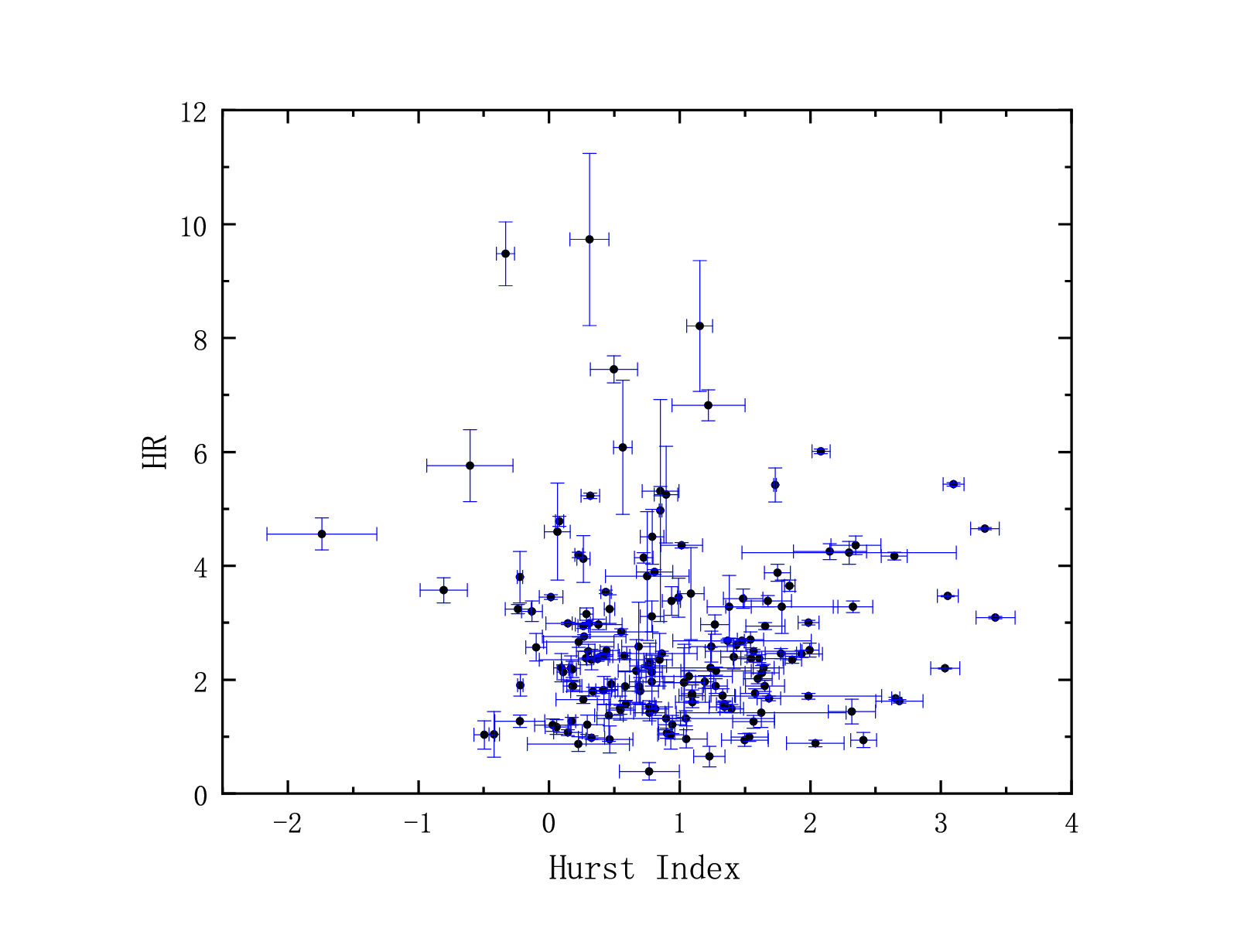}
    \caption{Scatter plot for $HR$ and Hurst index. For this parameter, the AIC/BIC model comparison favors the null (constant) model over the linear model; therefore, no regression line is shown.} The description of each parameter is in Section \ref{subsec:phys_para}.
    \label{fig:HR}
\end{figure}

For the spectral/flux-related parameters $\alpha$, $\beta$, $\log E_{\rm peak}$, $\log F_g$, $\log F_{pk}$, and $HR$, the corresponding scatter plots are shown in Figs.~\ref{fig:alpha}--\ref{fig:HR}. The corresponding Pearson's $r$ values are close to zero and the information criteria do not favor adding a slope (Table~\ref{tab:pearson}), indicating that the data provide no statistical evidence for a linear dependence on $H$ within our framework. Therefore, to avoid a potentially misleading emphasis on an unsupported trend, we do not list the regression relations for these parameters. For completeness, we still provide the best-fit slope and intercept, as well as Pearson's $r$, reduced $\chi^2$, and $\Delta$AIC/$\Delta$BIC in Table~\ref{tab:pearson}, enabling readers to directly compare the linear and constant models.

\begin{figure}
    \centering
    \includegraphics[width=0.7\linewidth]{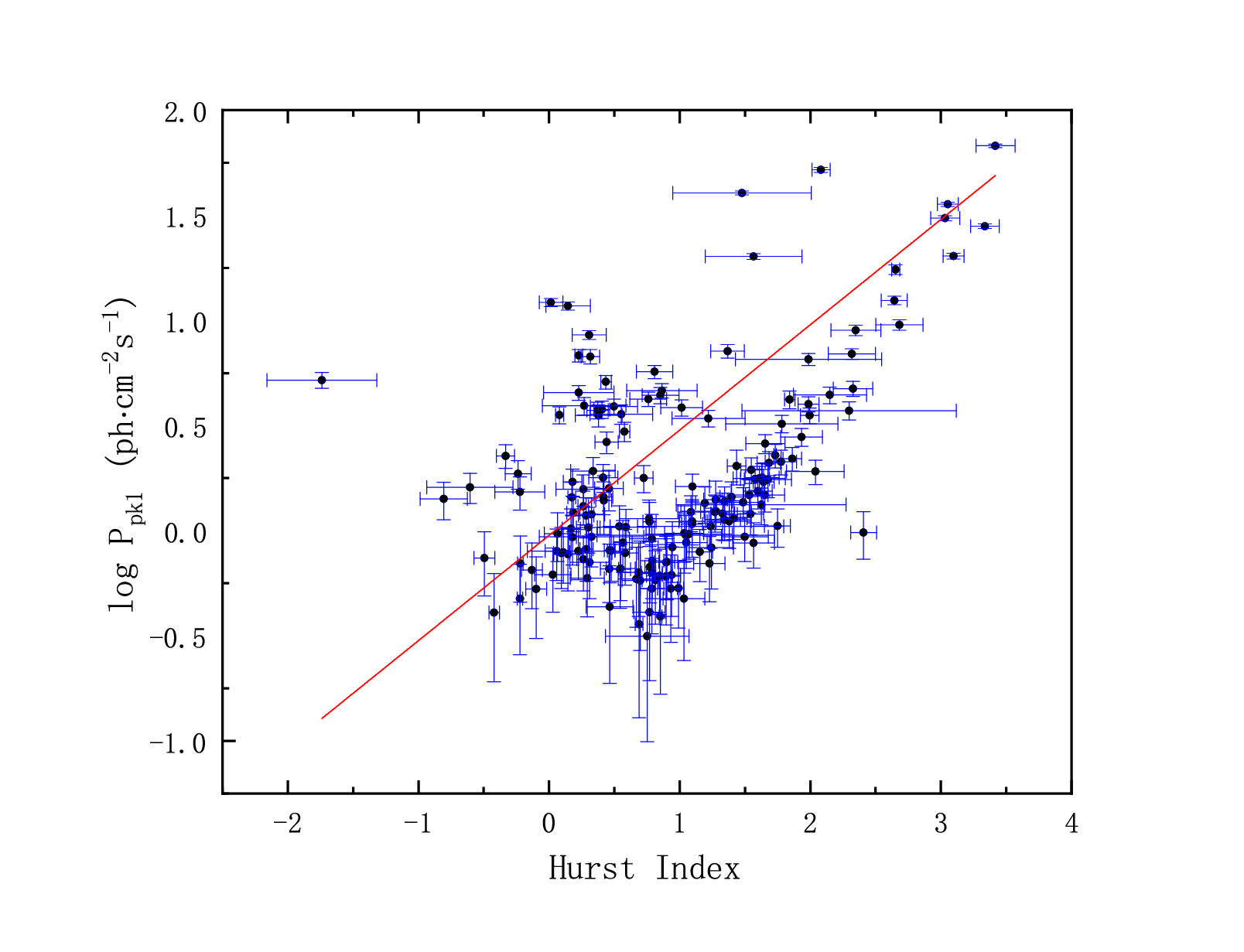}
    \caption{Scatter plot for $\log{P_{pk1}}$ and Hurst index. The solid line is our fit result. The relation for the solid line is $ \log{P_{pk1}}=(0.50\pm0.04)\times H +(-0.02\pm0.06)$. The description of each parameter is in Section \ref{subsec:phys_para}.}
    \label{fig:P_pk_1}
\end{figure}

The correlation between $\log{P_{pk1}}$ and $H$ is
\begin{equation}\label{eq:P_pk1}
    \log{P_{pk1}}=(0.50\pm0.04)\times H +(-0.02\pm0.06),
\end{equation}
where $P_{pk1}$ is the peak photon flux in the time bin $64$ ms of $10-1000$ keV and is in units of ph cm$^{-2}$s$^{-1}$. The Pearson's $r$ is $0.48^{+0.004}_{-0.05}$, the reduced $\chi^2$ is $28.49$. The $\Delta AIC$ and $\Delta BIC$ are 41.76 and 38.67, respectively. The scatter plot is in Fig. \ref{fig:P_pk_1}.

\begin{figure}
    \centering
    \includegraphics[width=0.7\linewidth]{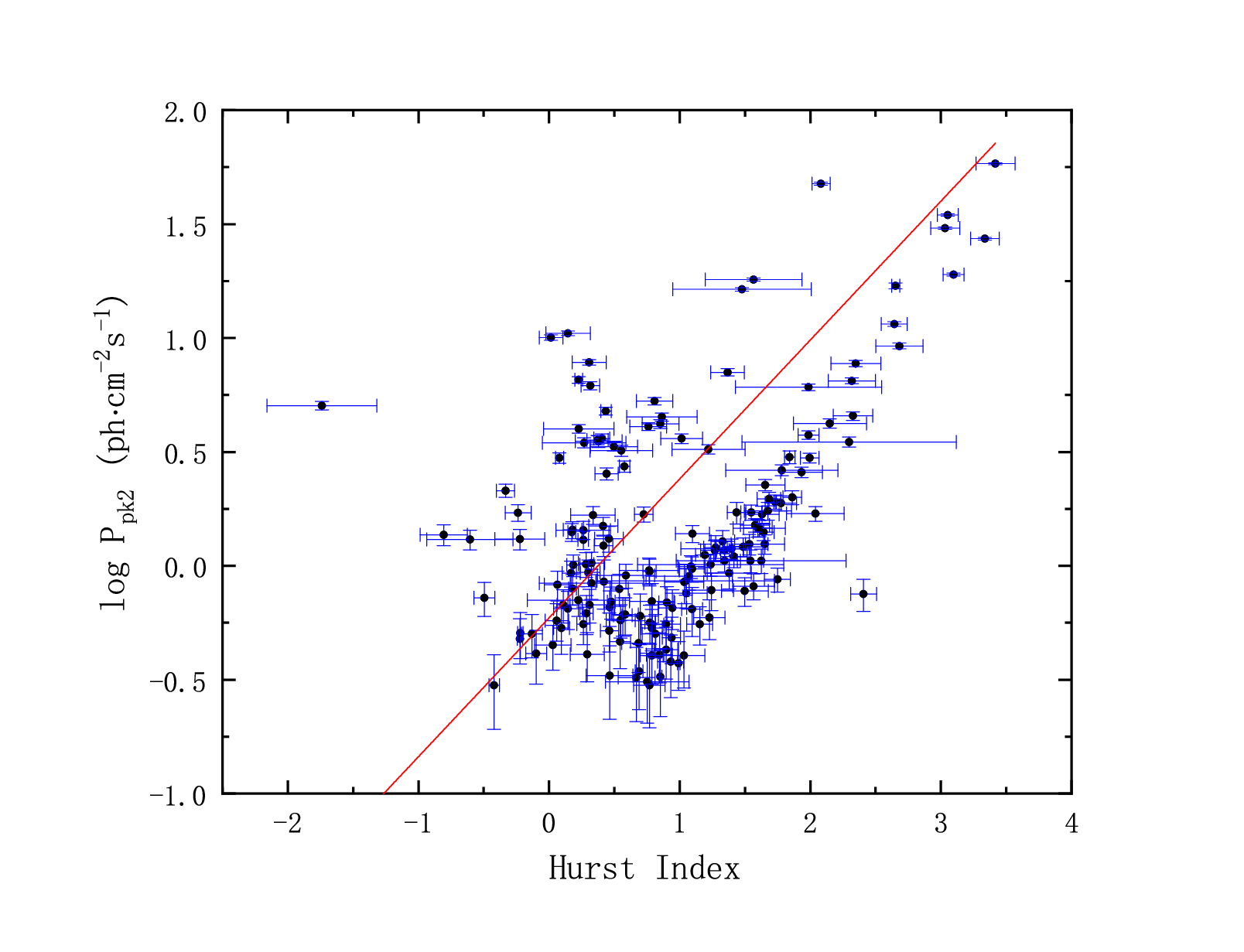}
    \caption{Scatter plot for $\log{P_{pk2}}$ and Hurst index. The solid line is our fit result. The relation for the solid line is $\log{P_{pk2}}=(0.61\pm0.05)\times H +(-0.23\pm0.07)$. The description of each parameter is in Section \ref{subsec:phys_para}.}
    \label{fig:P_pk_2}
\end{figure}

The correlation between $\log{P_{pk2}}$ and $H$ is
\begin{equation}\label{eq:P_pk2}
    \log{P_{pk2}}=(0.61\pm0.05)\times H +(-0.23\pm0.07),
\end{equation}
where $P_{pk2}$ is the peak photon flux in the time bin $256$ ms of $10-1000$ keV and is in units of ph cm$^{-2}$s$^{-1}$. The Pearson's $r$ is $0.48^{+0.01}_{-0.03}$, the reduced $\chi^2$ is $51.92$. The $\Delta AIC$ and $\Delta BIC$ are 41.55 and 38.45, respectively. The scatter plot is in Fig. \ref{fig:P_pk_2}.

\begin{figure}
    \centering
    \includegraphics[width=0.7\linewidth]{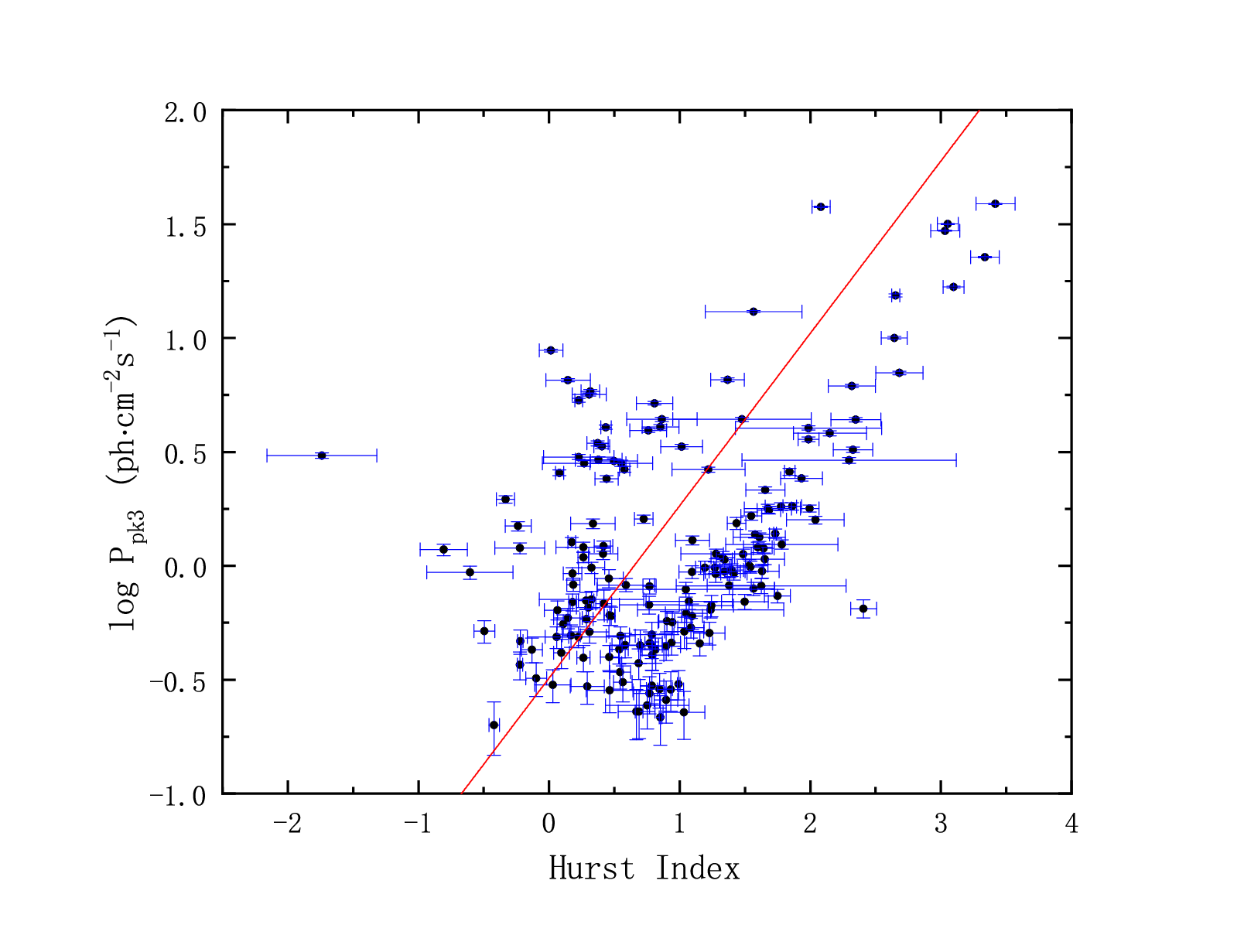}
    \caption{Scatter plot for $\log{P_{pk3}}$ and Hurst index. The solid line is our fit result. The relation for the solid line is $\log{P_{pk3}}=(0.76\pm0.07)\times H +(-0.49\pm0.09)$. The description of each parameter is in Section \ref{subsec:phys_para}.}
    \label{fig:P_pk_3}
\end{figure}

The correlation between $\log{P_{pk3}}$ and $H$ is
\begin{equation}\label{eq:P_pk3}
    \log{P_{pk3}}=(0.76\pm0.07)\times H +(-0.49\pm0.09),
\end{equation}
where $P_{pk3}$ is the peak photon flux in the time bin $1024$ ms of $10-1000$ keV and is in units of ph cm$^{-2}$s$^{-1}$. The Pearson's $r$ is $0.49^{+0.002}_{-0.03}$, the reduced $\chi^2$ is $72.54$. The $\Delta AIC$ and $\Delta BIC$ are 41.92 and 38.82, respectively. The scatter plot is in Fig. \ref{fig:P_pk_3}.

\begin{figure}
    \centering
    \includegraphics[width=0.7\linewidth]{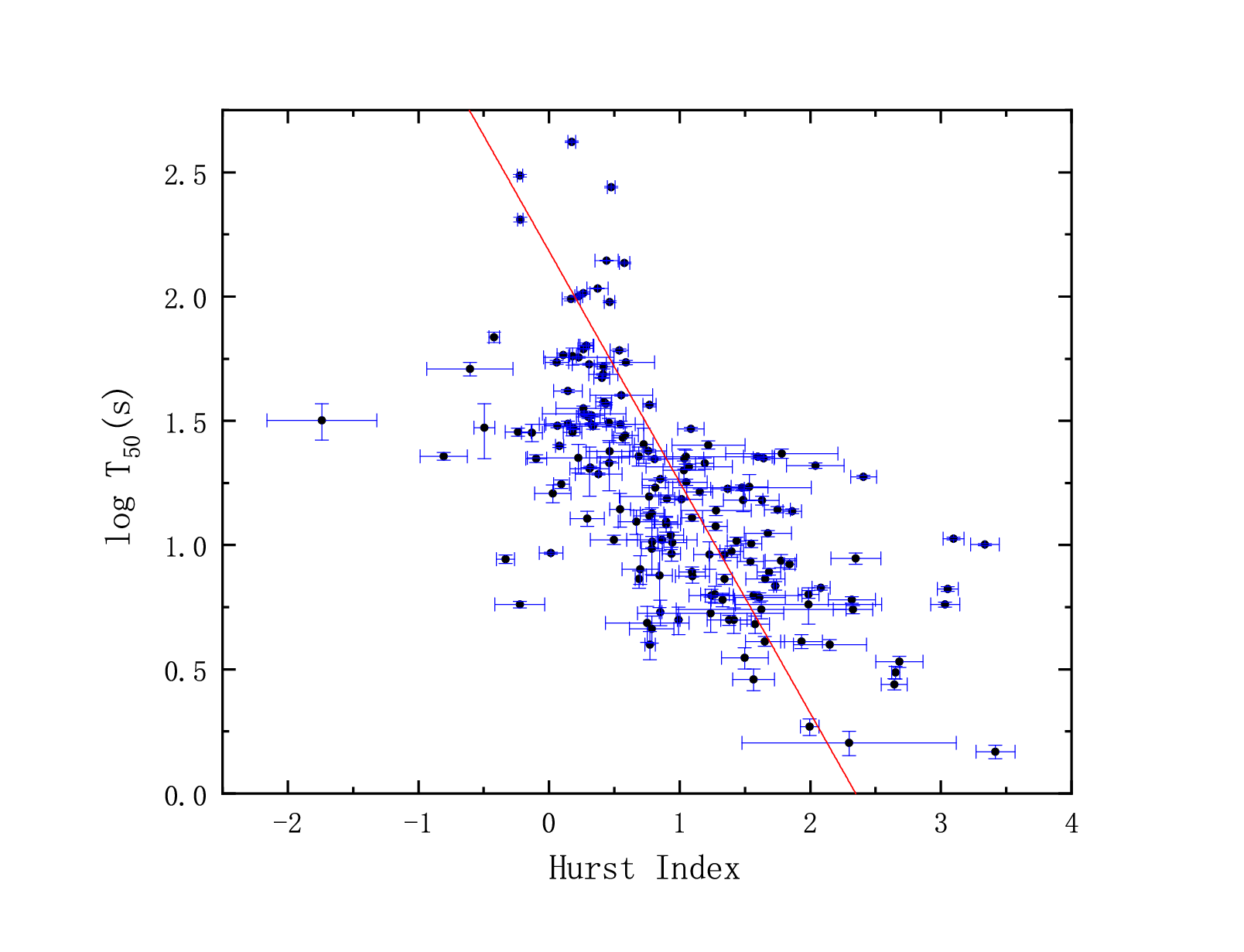}
    \caption{Scatter plot for $\log{T_{50}}$ and Hurst index. The solid line is our fit result. The relation for the solid line is $\log{T_{50}}=(-0.93\pm0.06)\times H +(2.18\pm0.05)$. The description of each parameter is in Section \ref{subsec:phys_para}.}
    \label{fig:T_50}
\end{figure}

The correlation between $\log{T_{50}}$ and $H$ is
\begin{equation}\label{eq:T_50}
    \log{T_{50}}=(-0.93\pm0.06)\times H +(2.18\pm0.05),
\end{equation}
where $T_{50}$ is the burst duration containing $25\%-75\%$ of total counts and is in units of s. The Pearson's $r$ is $-0.64^{+0.03}_{-0.002}$, the reduced $\chi^2$ is $54.23$. The $\Delta AIC$ and $\Delta BIC$ are 82.94 and 79.85, respectively. The scatter plot is in Fig. \ref{fig:T_50}.

\begin{figure}
    \centering
    \includegraphics[width=0.7\linewidth]{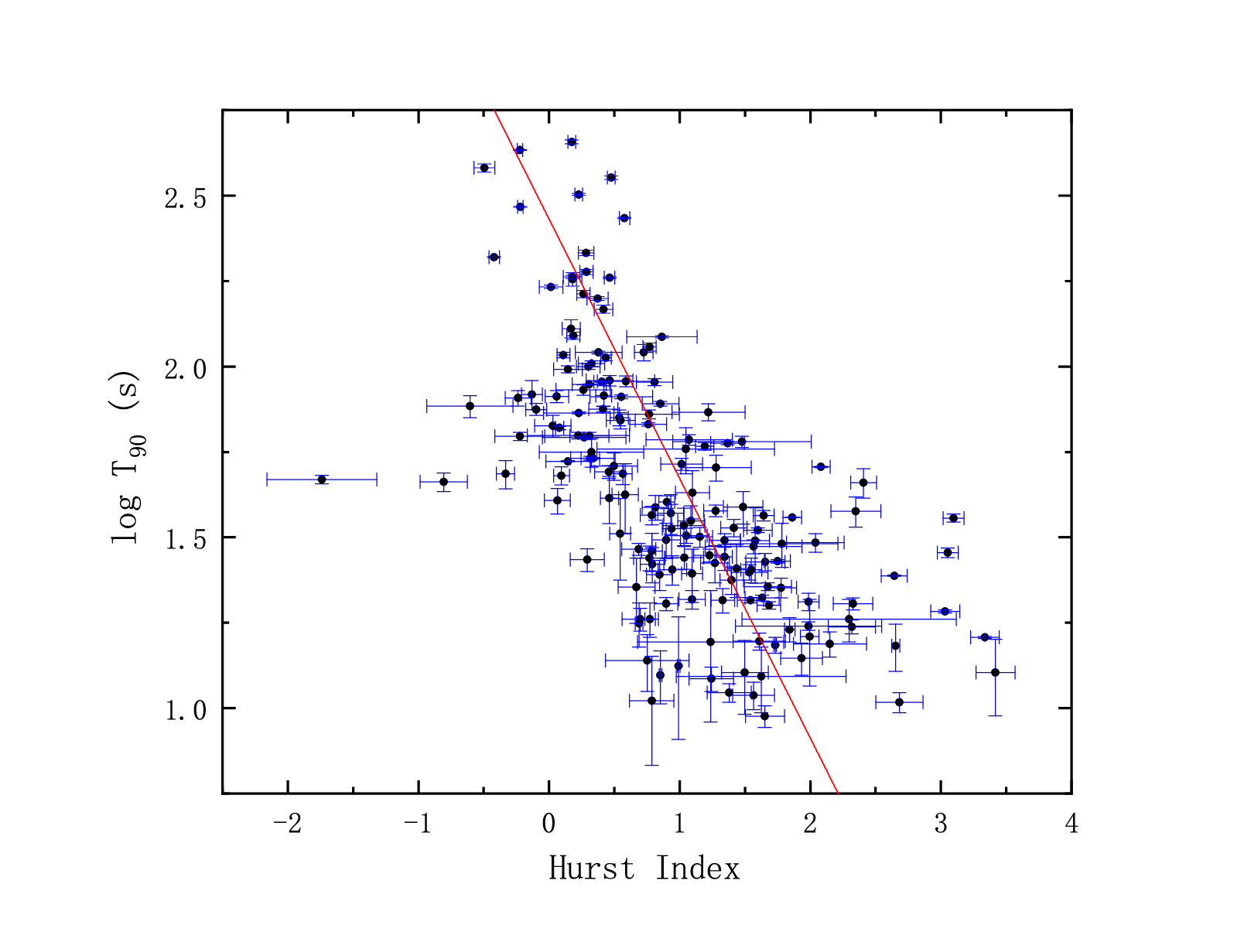}
    \caption{Scatter plot for $\log{T_{90}}$ and Hurst index. The solid line is our fit result. The relation for the solid line is $\log{T_{90}}=(-0.76\pm0.05)\times H +(2.43\pm0.04)$. The description of each parameter is in Section \ref{subsec:phys_para}.}
    \label{fig:T_90}
\end{figure}

The correlation between $\log{T_{90}}$ and $H$ is
\begin{equation}\label{eq:T_90}
    \log{T_{90}}=(-0.76\pm0.05)\times H +(2.43\pm0.04),
\end{equation}
where $T_{90}$ is the burst duration containing $5\%-95\%$ of total counts and is in units of s. The Pearson's $r$ is $-0.63^{+0.03}_{-0.004}$, the reduced $\chi^2$ is $41.82$. The $\Delta AIC$ and $\Delta BIC$ are 80.59 and 77.51, respectively. The scatter plot is in Fig. \ref{fig:T_90}.

\begin{figure}
    \centering
    \includegraphics[width=0.7\linewidth]{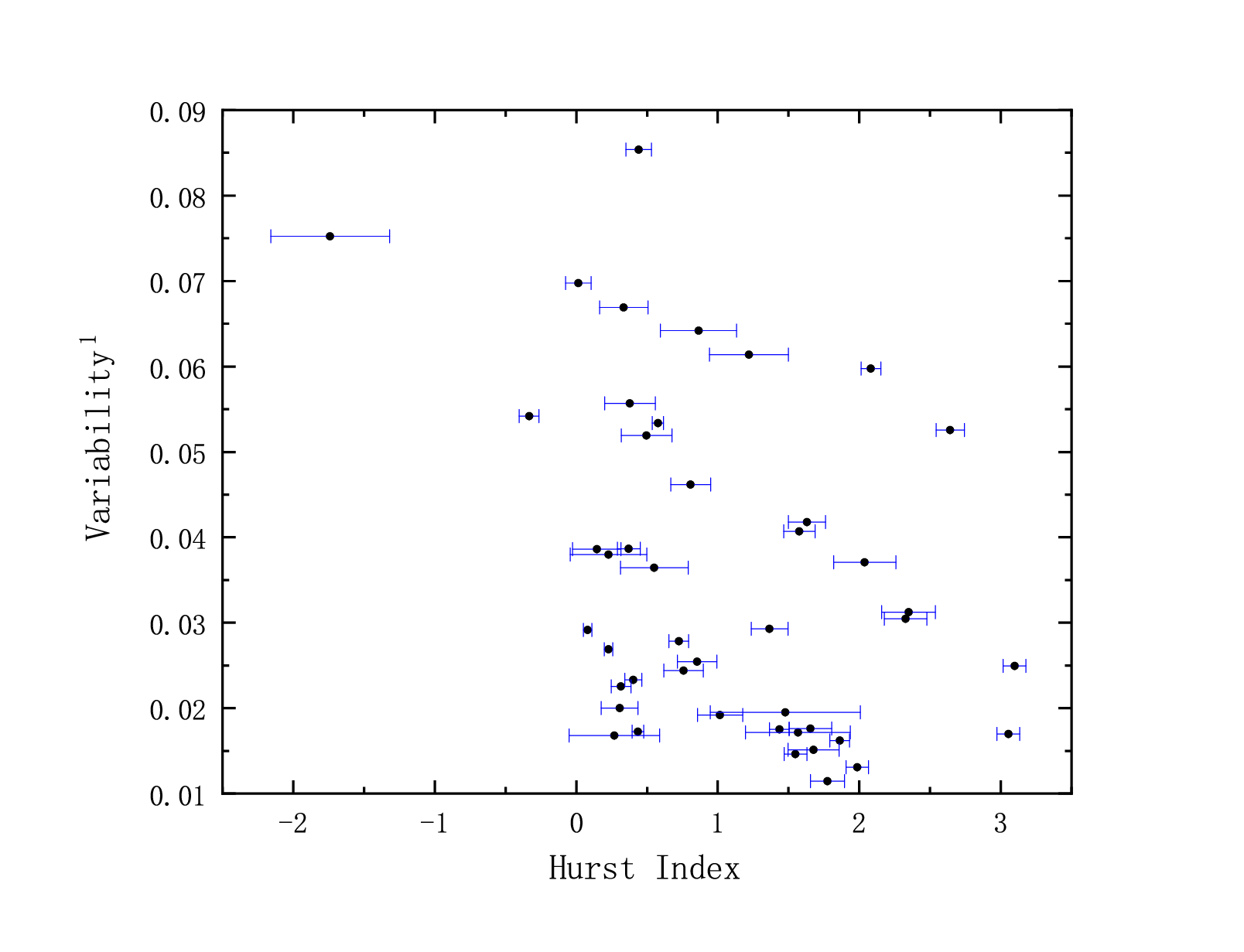}
    \caption{Scatter plot for $V$ and Hurst index. The description of each parameter is in Section \ref{subsec:phys_para}.}
    \label{fig:Variability_1}
\end{figure}

We further examine the light curve variability index $V$ defined by \citet{2000astro.ph..4176F}. 
The association between $V$ and $H$ is weak-to-moderate ($r=-0.39$), and we therefore do not emphasize a linear regression relation; for completeness, the corresponding regression coefficients are listed in Table~\ref{tab:pearson}. We do not report AIC/BIC or an uncertainty on Pearson's $r$ for $V$ because the variability index is not accompanied by a well-defined 1$\sigma$ measurement uncertainty in the data; without uncertainties for $V$, our likelihood-based AIC/BIC and MC error propagation of $r$ are not applicable in a consistent way. The scatter plot is shown in Fig. \ref{fig:Variability_1}.

\begin{table}[]
    \centering
    \resizebox{\linewidth}{!}{
    \begin{tabular}{ccccccc}
    \hline
    \hline
    Parameters vs. $H$ & Slope & Intercept &Pearson's $r$& Reduced $\chi^2$&$\Delta AIC$&$\Delta BIC$\\
     \hline
     $\alpha$ & $-0.48\pm0.05$ & $1.83\pm0.06$ &$-0.0033^{+0.03}_{-0.03}$&37.41 & -1.78&-4.88\\
     $\beta$ & $0.12\pm0.03$ & $2.23\pm0.06$ &$0.02^{+0.07}_{-0.09}$&5.72 &-337.55 &-340.11\\
     $\log{E_{peak}}$& $0.35\pm0.04$ & $1.83\pm0.06$ &$0.09^{+0.02}_{-0.04}$&55.84 & -0.69&-3.66\\
     $\log{F_g}$& $1.43\pm0.24$ & $-0.39\pm0.27$ &$0.13^{+0.01}_{-0.03}$&117.76 & 0.74&-2.34\\
     $\log{F_{pk}}$ & $-0.19\pm0.12$ & $-0.07\pm0.18$ &$-0.24^{+0.07}_{-0.05}$&29.13 &-1.41 &-1.97\\
     $HR$& $2.92\pm0.58$ & $-0.04\pm0.52$ &$-0.01^{+0.02}_{-0.02}$&99.90 & -1.97&-5.05\\
     $\log{P_{pk1}}$& $0.50\pm0.04$ & $-0.02\pm0.06$ &$0.48^{+0.004}_{-0.05}$&28.49 &41.76 &38.67\\
     $\log{P_{pk2}}$& $0.61\pm0.05$ & $-0.23\pm0.07$ &$0.48^{+0.01}_{-0.03}$&51.92 &41.55 &38.45\\
     $\log{P_{pk3}}$& $0.76\pm0.07$  & $-0.49\pm0.09$ &$0.49^{+0.002}_{-0.03}$&72.54 & 41.92&38.82\\
     $\log{T_{50}}$& $-0.93\pm0.06$  & $2.18\pm0.05$ &$-0.64^{+0.03}_{-0.002}$& 54.23& 82.94&79.85\\
     $\log{T_{90}}$& $-0.76\pm0.05$ &  $2.43\pm0.04$ &$-0.63^{+0.03}_{-0.004}$&41.82 &80.59 &77.51\\
     $\rm{V}$& $-0.01\pm0.16$ & $0.04\pm0.22$  &$-0.39$& --& --&--\\
      \hline
    \end{tabular}
    }
    \caption{Correlation coefficient results.}
    \label{tab:pearson}
\end{table}

The correlation analysis between the Hurst index ($H$) and the 12 GRB observables is summarized in Table~\ref{tab:pearson}. The complete set of best-fit linear relations is given in Eqs.~(\ref{eq:P_pk1})--(\ref{eq:T_90}), and the corresponding scatter plots are shown in Figs.~\ref{fig:alpha}--\ref{fig:Variability_1}.
In addition to the Pearson coefficient, we report the reduced $\chi^2$ from the linear fits and the information-criterion differences ($\Delta\mathrm{AIC}$ and $\Delta\mathrm{BIC}$) comparing the linear model against a null (no-trend) model (Table~\ref{tab:pearson}).

Overall, the strongest (anti-)correlations are found for the temporal parameters $T_{50}$ and $T_{90}$, with Pearson coefficients $r=-0.64^{+0.03}_{-0.002}$ and $r=-0.63^{+0.03}_{-0.004}$, respectively. Consistent with these coefficients, both $\log T_{50}$ and $\log T_{90}$ yield the largest positive $\Delta\mathrm{AIC}$ and $\Delta\mathrm{BIC}$ among all parameters, indicating that a linear trend is strongly preferred over the null model for these duration-related observables.

For the flux-related quantities, we find clear positive associations between $H$ and the peak photon flux proxies: $\log P_{pk1}$, $\log P_{pk2}$, and $\log P_{pk3}$ all show moderate positive correlations with $r\simeq 0.48$--$0.49$ (Table~\ref{tab:pearson}). The corresponding $\Delta\mathrm{AIC}$ and $\Delta\mathrm{BIC}$ are also positive and similar for the three time bins, supporting the presence of a linear trend. By contrast, for both $\log F_g$ and $\log F_{pk}$ we find no evidence that a linear trend with $H$ is required by the data.
The Pearson coefficients are $r=0.13^{+0.01}_{-0.03}$ for $\log F_g$ and $r=-0.24^{+0.07}_{-0.05}$ for $\log F_{pk}$, and the information-criterion comparison ($\Delta\mathrm{AIC}$ and $\Delta\mathrm{BIC}$) favors the null model over the linear model in both cases.
Accordingly, we do not emphasize linear regression relations; for completeness, the linear-fit coefficients together with the corresponding $\Delta\mathrm{AIC}$ and $\Delta\mathrm{BIC}$ are listed in Table~\ref{tab:pearson}.

For the spectral- and hardness-related parameters, the data show no compelling evidence for a linear dependence on $H$. Consistent with the near-zero Pearson coefficients, information-criterion model comparison further indicates that the null model is preferred over the linear model for these parameters; accordingly, we do not display linear regression lines in the corresponding figures.
In particular, the Pearson coefficient for $\alpha$ is consistent with zero, with $r=-0.0033^{+0.03}_{-0.03}$, and $HR$ is also consistent with no correlation, with $r=-0.01^{+0.02}_{-0.02}$. 
The variability index $V$ shows a negative Pearson coefficient $r=-0.39$; however, given the absence of a well-defined $1\sigma$ measurement uncertainty for $V$ in the catalog, we do not apply our likelihood-based $\Delta \mathrm{AIC}/\Delta \mathrm{BIC}$ model comparison or Monte-Carlo uncertainty propagation for $r$ to this parameter in order to maintain statistical consistency.

Finally, we note that the reduced $\chi^2$ values for several relations are substantially larger than unity (Table~\ref{tab:pearson}), indicating that the observed scatter is not fully explained by the quoted measurement errors under a simple linear description. We therefore interpret the Pearson coefficient and $\Delta\mathrm{AIC}/\Delta\mathrm{BIC}$ as complementary diagnostics: $r$ characterizes the monotonic association strength, while the goodness-of-fit and information criteria quantify whether a linear model provides an adequate and parsimonious description of the data.

\section{Conclusions and Discussion} \label{sec:discussion_conclusions}
In this work, we applied detrended fluctuation analysis (DFA) to the prompt-emission light curves of 163 long-duration BATSE GRBs and extracted the Hurst index $H$ as a quantitative descriptor of the scaling behavior of the fluctuation function. We then examined how $H$ relates to twelve commonly used GRB observables compiled by \citet{Wang2020}, and summarized the correlation coefficients and linear-fit diagnostics in Table~\ref{tab:pearson}.

Our most robust empirical result is a strong anti-correlation between $H$ and the duration-related parameters: $\log T_{50}$ and $\log T_{90}$ yield Pearson coefficients of $r=-0.64^{+0.03}_{-0.002}$ and $r=-0.63^{+0.03}_{-0.004}$, respectively, and both relations are strongly preferred over a null (no-trend) model according to $\Delta\mathrm{AIC}$ and $\Delta\mathrm{BIC}$ (Table~\ref{tab:pearson}).
This indicates that GRBs with longer durations tend to exhibit smaller DFA-derived $H$, i.e., weaker persistence in the scaling behavior of their prompt light curves. One plausible interpretation is that longer bursts may involve a larger number of emission episodes and/or a more heterogeneous superposition of variability components, which reduces the apparent long-range persistence captured by a single global DFA slope.

In the flux-related group, we find moderate positive correlations between $H$ and the peak photon flux proxies, with $r\simeq 0.48$--$0.49$ for $\log P_{pk1}$, $\log P_{pk2}$, and $\log P_{pk3}$, and consistently positive $\Delta\mathrm{AIC}$ and $\Delta\mathrm{BIC}$ favoring a linear trend (Table~\ref{tab:pearson}).
This suggests that bursts with more intense prompt emission (as traced by peak photon flux) tend to display larger $H$ values. Notably, within our sample, the correlation strengths for $P_{pk1}$--$P_{pk3}$ are comparable rather than showing a clear monotonic weakening from 64\,ms to 1024\,ms, indicating that the association between $H$ and peak intensity is not confined to a single peak-flux timescale in the Wang~et~al. compilation.

By contrast, the spectral and hardness-related quantities, our analysis does not support a statistically meaningful linear dependence on $H$. 
The Pearson coefficients are consistent with no correlation or are at most marginal in magnitude: 
$\alpha$ ($r=-0.0033^{+0.03}_{-0.03}$), $\beta$ ($r=0.02^{+0.07}_{-0.09}$), $\log E_{peak}$ ($r=0.09^{+0.02}_{-0.04}$), and $HR$ ($r=-0.01^{+0.02}_{-0.02}$). 
Consistently, the information-criterion comparison ($\Delta\mathrm{AIC}$ and $\Delta\mathrm{BIC}$; Table~\ref{tab:pearson}) favors the null model over the linear model for these parameters.
Accordingly, we do not claim a strong linkage between the DFA scaling behavior and the Band-function spectral indices based on the present BATSE sample and analysis choices. The variability index is treated separately because we do not propagate uncertainty on $r$ nor compute AIC/BIC for $V$ due to the lack of a well-defined 1$\sigma$ measurement uncertainty for $V$ in the catalog.

Several caveats should be considered when interpreting these trends. First, many reduced $\chi^2$ values in Table~\ref{tab:pearson} are substantially larger than unity, implying that a simple linear relation does not fully account for the scatter under the quoted measurement errors; intrinsic dispersion and/or unmodeled heterogeneity across the GRB population is likely important. Second, our DFA implementation adopts uniform binning and a single detrending order, and we fit a single slope over all adopted scales; GRBs with scale-dependent behavior (e.g., crossovers) may not be optimally characterized by a single global $H$. Third, all results are based on BATSE DISCSC light curves and the parameter definitions adopted by \citet{Wang2020}; correlations may change with broader energy coverage, different temporal sampling, or alternative catalog constructions. Finally, for a small subset of bursts with truncated coverage ($t_{\rm end}^{\rm data}<T_{90}$) and/or collapsed scale range ($s_{\max}<s_{\min}$), the DFA fit is less constrained and thus expected to carry larger statistical uncertainties, reducing their leverage in sample-level correlations.

From Figs. \ref{fig:F_g} -- \ref{fig:P_pk_3}, one can see a weak signature indicating that the sample can be classified into two classes in the Hurst index and flux plane. This phenomenon appears in all fluences, peak flux, and peak photon flux. It is similar to the Amati relation \citep{2002A&A...390...81A} and the Ghirlanda relation \citep{2004ApJ...616..331G}, which can be used to distinguish between the collapsar origin and the binary merging origin. However, as the samples are all long GRBs, it should not be the case for differentiating between long and short GRBs. This may provide another criterion for the classification of GRB sub-classes. These sub-classes might be used for those bursts with redshift measurements in future work. Then the physical quantities, such as luminosity and total energy, can be included, and the underlying origin of the sub-classes might be discovered.

Future work can extend this study by applying the same pipeline to broader-band datasets (e.g., Fermi/GBM), exploring time-resolved or scale-dependent DFA slopes within individual bursts, and comparing the observed $H$--duration and $H$--peak-flux trends with predictions from physically motivated prompt-emission simulations. In addition, we will quantify these potential limitations via robustness checks that repeat the correlation tests under stricter thresholds on the minimum time-series length and the minimum number of usable scales.

In summary, the DFA-derived Hurst index $H$ shows its clearest connections to prompt-emission timescales (durations) and intensity indicators (peak photon fluxes), while the standard spectral parameters provide no comparable evidence for a linear dependence on $H$ in the present sample. These empirical relations provide a quantitative link between the statistical structure of prompt light-curve variability and a subset of key GRB observables and motivate further multi-instrument and time-resolved investigations.

\section*{Acknowledgments}
We thank the anonymous reviewer for his/her comments which enhanced this work enormously. We also thank Shuang-Xi Yi for useful comments and Cheng-Jie Sun for useful discussions. This work was supported by the National SKA Program of China (2022SKA0130100). The computation was performed on the HPC Platform of Huazhong University of Science and Technology.

\appendix
\section{List of Hurst Indices}\label{appendix}

\begin{center}
\begin{longtable}{ccc}
\caption{List of Hurst Indices.} \label{tab:list_HI} \\
\hline
\hline
Trigger number & GRB name & Hurst index \\
\hline
\endfirsthead

\multicolumn{3}{c}{{\tablename\ \thetable{} -- Continued from previous page}} \\
\hline
\hline
Trigger number & GRB name & Hurst index \\
\hline
\endhead

\hline
\multicolumn{3}{r}{{Continued on next page}} \\
\endfoot

\hline
\hline
\endlastfoot

107 & GRB 910423 & $-0.42\pm 0.04$ \\
{109} & GRB 910425A & $0.40\pm 0.06$ \\
{110} & GRB 910425B & $-0.22\pm 0.02$ \\
{111} & GRB 910426 & $0.14\pm 0.11$ \\
{114} & GRB 910427 & $-0.50\pm 0.08$ \\
121 & GRB 910429 & $0.42\pm 0.11$ \\
130 & GRB 910430 & $0.27\pm 0.32$ \\
133 & GRB 910501 & $0.46\pm 0.04$ \\
143 & GRB 910503 & $2.08\pm 0.07$ \\
148 & GRB 910505 & $0.17\pm 0.03$ \\
160 & GRB 910507 & $1.84\pm 0.04$ \\
171 & GRB 910509 & $1.10\pm 0.08$ \\
204 & GRB 910517{B} & $0.31\pm 0.15$ \\
211 & GRB 910518{B} & $-0.22\pm 0.02$ \\
214 & GRB 910521{B} & $2.41\pm 0.10$ \\
219 & GRB 910522 & $1.57\pm 0.37$ \\
222 & GRB 910523 & $0.23\pm 0.27$ \\
223 & GRB 910523B & $0.69\pm 0.13$ \\
226 & GRB 910525 & $0.26\pm 0.04$ \\
228 & GRB 910526{B} & $1.62\pm 0.65$ \\
235 & GRB 910528 & $0.79\pm 0.09$ \\
237 & GRB 910529 & $0.06\pm 0.10$ \\
249 & GRB 910601 & $3.05\pm 0.08$ \\
257 & GRB 910602 & $-0.24\pm 0.10$ \\
288 & GRB 910607{B} & $0.46\pm 0.07$ \\
332 & GRB 910612 & $1.28\pm 0.27$ \\
351 & GRB 910614{B} & $0.42\pm 0.07$ \\
394 & GRB 910619 & $0.44\pm 0.04$ \\
398 & GRB 910620 & $1.44\pm 0.07$ \\
404 & GRB 910621B & $0.19\pm 0.05$ \\
408 & GRB 910621 & $0.26\pm 0.21$ \\
414 & GRB 910622 & $1.24\pm 0.56$ \\
451 & GRB 910627 & $2.65\pm 0.03$ \\
465 & GRB 910629B & $0.75\pm 0.32$ \\
467 & GRB 910629 & $2.35\pm 0.19$ \\
469 & GRB 910630 & $2.15\pm 0.28$ \\
472 & GRB 910701 & $0.42\pm 0.06$ \\
473 & GRB 910702{B} & $0.90\pm 0.11$ \\
501 & GRB 910708 & $1.41\pm 0.10$ \\
503 & GRB 910709 & $-1.74\pm 0.42$ \\
516 & GRB 910712 & $1.50\pm 0.18$ \\
540 & GRB 910715 & $1.05\pm 0.16$ \\
548 & GRB 910718{B} & $1.86\pm 0.07$ \\
559 & GRB 910721B & $1.03\pm 0.16$ \\
563 & GRB 910721 & $1.78\pm 0.12$ \\
577 & GRB 910725{B} & $0.18\pm 0.03$ \\
591 & GRB 910730B & $0.33\pm 0.40$ \\
594 & GRB 910730 & $1.75\pm 0.10$ \\
606 & GRB 910802{B} & $1.54\pm 0.10$ \\
630 & GRB 910805 & $0.70\pm 0.14$ \\
647 & GRB 910807 & $1.37\pm 0.13$ \\
658 & GRB 910809C & $1.33\pm 0.09$ \\
659 & GRB 910809B & $0.32\pm 0.10$ \\
660 & GRB 910809 & $2.33\pm 0.15$ \\
673 & GRB 910813 & $0.77\pm 0.12$ \\
676 & GRB 910814C & $0.85\pm 0.14$ \\
678 & GRB 910814A & $0.32\pm 0.07$ \\
680 & GRB 910815 & $1.57\pm 0.16$ \\
685 & GRB 910816B & $0.77\pm 0.23$ \\
686 & GRB 910816 & $1.38\pm 0.17$ \\
692 & GRB 910818 & $-0.22\pm 0.19$ \\
704 & GRB 910821 & $-0.81\pm 0.18$ \\
717 & GRB 910823 & $0.93\pm 0.10$ \\
741 & GRB 910829 & $0.85\pm 0.01$ \\
753 & GRB 910903 & $0.77\pm 0.04$ \\
761 & GRB 910905 & $0.55\pm 0.24$ \\
764 & GRB 910907 & $1.10\pm 0.10$ \\
773 & GRB 910908 & $1.07\pm 0.33$ \\
795 & GRB 910914 & $1.19\pm 0.07$ \\
815 & GRB 910923 & $0.94\pm 0.11$ \\
816 & GRB 910925 & $0.46\pm 0.18$ \\
820 & GRB 910926 & $1.15\pm 0.10$ \\
824 & GRB 910926B & $0.90\pm 0.06$ \\
825 & GRB 910927B & $-0.10\pm 0.08$ \\
829 & GRB 910927 & $2.32\pm 0.18$ \\
840 & GRB 910930B & $-0.61\pm 0.33$ \\
841 & GRB 910930 & $1.99\pm 0.56$ \\
869 & GRB 911005 & $0.38\pm 0.18$ \\
907 & GRB 911016 & $0.37\pm 0.08$ \\
927 & GRB 911024 & $1.23\pm 0.12$ \\
938 & GRB 911026 & $1.34\pm 0.12$ \\
946 & GRB 911027{B} & $0.56\pm 0.07$ \\
973 & GRB 911031{A} & $0.81\pm 0.14$ \\
1009 & GRB 911106{B} & $0.48\pm 0.03$ \\
1036 & GRB 911110 & $0.30\pm 0.09$ \\
1039 & GRB 911111C & $1.10\pm 0.13$ \\
1042 & GRB 911111 & $1.53\pm 0.14$ \\
1046 & GRB 911111B & $0.77\pm 0.05$ \\
1085 & GRB 911118A & $3.03\pm 0.11$ \\
1086 & GRB 911118B & $0.67\pm 0.14$ \\
1087 & GRB 911119{B} & $1.03\pm 0.15$ \\
1156 & GRB 911209C & $0.29\pm 0.05$ \\
1157 & GRB 911209 & $0.01\pm 0.09$ \\
1159 & GRB 911210 & $2.30\pm 0.82$ \\
1167 & GRB 911213 & $0.84\pm 0.10$ \\
1192 & GRB 911217{B} & $0.58\pm 0.10$ \\
1196 & GRB 911219B & $0.26\pm 0.05$ \\
1197 & GRB 911219 & $1.40\pm 0.09$ \\
1200 & GRB 911221 & $1.34\pm 0.06$ \\
1213 & GRB 911224{C} & $0.90\pm 0.09$ \\
1218 & GRB 911225 & $1.65\pm 0.15$ \\
1235 & GRB 911227 & $0.44\pm 0.09$ \\
1244 & GRB 911228 & $1.05\pm 0.68$ \\
1279 & GRB 920105 & $1.27\pm 0.09$ \\
1288 & GRB 920110A & $0.23\pm 0.03$ \\
1291 & GRB 920110B & $1.73\pm 0.01$ \\
1384 & GRB 920210C & $0.46\pm 0.11$ \\
1385 & GRB 920210 & $1.02\pm 0.16$ \\
1390 & GRB 920212 & $0.69\pm 0.03$ \\
1396 & GRB 920214 & $1.63\pm 0.13$ \\
1406 & GRB 920216 & $1.68\pm 0.09$ \\
1419 & GRB 920218 & $0.86\pm 0.27$ \\
1425 & GRB 920221 & $2.68\pm 0.18$ \\
1432 & GRB 920224 & $0.81\pm 0.10$ \\
1440 & GRB 920226 & $2.64\pm 0.10$ \\
1446 & GRB 920227C & $0.94\pm 0.07$ \\
1447 & GRB 920227B & $1.68\pm 0.18$ \\
1449 & GRB 920228 & $1.60\pm 0.11$ \\
1452 & GRB 920229{B} & $1.24\pm 0.17$ \\
1456 & GRB 920301 & $0.09\pm 0.06$ \\
1458 & GRB 920302{B} & $0.06\pm 0.09$ \\
1467 & GRB 920307 & $1.66\pm 0.15$ \\
1468 & GRB 920308{A} & $0.50\pm 0.18$ \\
1472 & GRB 920310 & $1.61\pm 0.20$ \\
1574 & GRB 920430 & $0.11\pm 0.05$ \\
1578 & GRB 920502 & $1.99\pm 0.08$ \\
1579 & GRB 920502C & $1.22\pm 0.28$ \\
1580 & GRB 920503 & $1.49\pm 0.15$ \\
1586 & GRB 920505 & $0.59\pm 0.22$ \\
1590 & GRB 920509 & $0.55\pm 0.11$ \\
1601 & GRB 920511A & $-0.33\pm 0.07$ \\
1604 & GRB 920511B & $0.79\pm 0.08$ \\
1606 & GRB 920513 & $0.31\pm 0.13$ \\
1609 & GRB 920517 & $3.42\pm 0.15$ \\
1611 & GRB 920518 & $1.09\pm 0.10$ \\
1614 & GRB 920520 & $0.28\pm 0.06$ \\
1623 & GRB 920524 & $0.08\pm 0.03$ \\
1625 & GRB 920525 & $3.34\pm 0.11$ \\
1626 & GRB 920525C & $1.78\pm 0.43$ \\
1628 & GRB 920526 & $1.64\pm 0.08$ \\
1642 & GRB 920609 & $-0.13\pm 0.08$ \\
1646 & GRB 920613 & $0.17\pm 0.07$ \\
1652 & GRB 920617 & $0.76\pm 0.14$ \\
1653 & GRB 920617C & $0.22\pm 0.39$ \\
1655 & GRB 920618 & $0.03\pm 0.14$ \\
1656 & GRB 920619 & $0.18\pm 0.07$ \\
1657 & GRB 920619B & $0.99\pm 0.01$ \\
1660 & GRB 920620 & $0.34\pm 0.17$ \\
1661 & GRB 920620C & $0.79\pm 0.09$ \\
1663 & GRB 920622 & $3.10\pm 0.08$ \\
1667 & GRB 920624 & $0.29\pm 0.13$ \\
1676 & GRB 920627 & $0.15\pm 0.17$ \\
1687 & GRB 920707 & $0.55\pm 0.08$ \\
1693 & GRB 920710 & $0.79\pm 0.17$ \\
1922 & GRB 920912 & $2.00\pm 0.07$ \\
1924 & GRB 920913 & $1.27\pm 0.11$ \\
1956 & GRB 920925 & $1.93\pm 0.16$ \\
1967 & GRB 921001 & $1.55\pm 0.08$ \\
1982 & GRB 921008 & $0.73\pm 0.07$ \\
1989 & GRB 921015 & $0.58\pm 0.04$ \\
1991 & GRB 921017 & $0.54\pm 0.07$ \\
1993 & GRB 921021 & $2.04\pm 0.22$ \\
1997 & GRB 921022 & $1.48\pm 0.53$ \\

\end{longtable}
\end{center}

\vspace{5mm}
\newcommand{\actaa}{Acta Astronomica}
\newcommand{\araa}{Annu.\ Rev.\ Astron.\ Astroph.}
\newcommand{\arnps}{Annu.\ Rev.\ Nucl.\ Part.\ Sci.}
\newcommand{\al}{Astron.\ Lett.}
\newcommand{\aap}{Astron.\ Astrophys.}
\newcommand{\apj}{Astrophys.\ J.}
\newcommand{\apjl}{Astrophys.\ J.}
\newcommand{\apjs}{Astrophys.\ J.\ Suppl.\ Ser.}
\newcommand{\epja}{Eur.\ Phys.\ J. A}
\newcommand{\iaucirc}{IAU Circ.}
\newcommand{\jaa}{J.\ Astron.\ Astrophys.}
\newcommand{\nar}{New Astronomy Reviews}
\newcommand{\nat}{Nature}
\newcommand{\mnras}{Mon.\ Not.\ R.\ Astron.\ Soc.}
\newcommand{\npa}{Nucl.\ Phys. A}
\newcommand{\pasp}{PASP}
\newcommand{\physrep}{Phys.\ Rep.}
\newcommand{\prc}{Phys.\ Rev. C}
\newcommand{\prd}{Phys.\ Rev. D}
\newcommand{\pre}{Phys.\ Rev. E}
\newcommand{\prl}{Phys.\ Rev.\ Lett.}
\newcommand{\ptp}{Prog.\ Theor.\ Phys.}
\newcommand{\ppnp}{Prog.\ Part.\ Nucl.\ Phys.}
\newcommand{\rmp}{Rev.\ Mod.\ Phys.}
\newcommand{\ssr}{Space Sci.\ Rev.}
\newcommand{\solphys}{Solar Phys.}

\bibliographystyle{elsarticle-harv}
\bibliography{ref}{}

@ARTICLE{1994PhRvE..49.1685P,
       author = {{Peng}, C. -K. and {Buldyrev}, S.~V. and {Havlin}, S. and {Simons}, M. and {Stanley}, H.~E. and {Goldberger}, A.~L.},
        title = "{Mosaic organization of DNA nucleotides}",
      journal = {\pre},
         year = 1994,
        month = feb,
       volume = {49},
       number = {2},
        pages = {1685-1689},
          doi = {10.1103/PhysRevE.49.1685},
       adsurl = {https://ui.adsabs.harvard.edu/abs/1994PhRvE..49.1685P}
}

@ARTICLE{1995Chaos...5...82P,
       author = {{Peng}, C. -K. and {Havlin}, Shlomo and {Stanley}, H. Eugene and {Goldberger}, Ary L.},
        title = "{Quantification of scaling exponents and crossover phenomena in nonstationary heartbeat time series}",
      journal = {Chaos},
         year = 1995,
        month = mar,
       volume = {5},
       number = {1},
        pages = {82-87},
          doi = {10.1063/1.166141},
       adsurl = {https://ui.adsabs.harvard.edu/abs/1995Chaos...5...82P}
}

@ARTICLE{2002PhyA..316...87K,
       author = {{Kantelhardt}, Jan W. and {Zschiegner}, Stephan A. and {Koscielny-Bunde}, Eva and {Havlin}, Shlomo and {Bunde}, Armin and {Stanley}, H. Eugene},
        title = "{Multifractal detrended fluctuation analysis of nonstationary time series}",
      journal = {Physica A Statistical Mechanics and its Applications},
         year = 2002,
        month = dec,
       volume = {316},
       number = {1},
        pages = {87-114},
          doi = {10.1016/S0378-4371(02)01383-3},
archivePrefix = {arXiv},
       eprint = {physics/0202070},
 primaryClass = {physics.data-an},
       adsurl = {https://ui.adsabs.harvard.edu/abs/2002PhyA..316...87K}
}

@ARTICLE{2006JSMTE..02..003S,
       author = {{Sadegh Movahed}, M. and {Jafari}, G.~R. and {Ghasemi}, F. and {Rahvar}, Sohrab and {Rahimi Tabar}, M. Reza},
        title = "{Multifractal detrended fluctuation analysis of sunspot time series}",
      journal = {Journal of Statistical Mechanics: Theory and Experiment},
         year = 2006,
        month = feb,
       volume = {2006},
       number = {2},
        pages = {02003},
          doi = {10.1088/1742-5468/2006/02/P02003},
archivePrefix = {arXiv},
       eprint = {physics/0508149},
 primaryClass = {physics.data-an},
       adsurl = {https://ui.adsabs.harvard.edu/abs/2006JSMTE..02..003S}
}

@ARTICLE{2009JSMTE..02..066H,
       author = {{Hu}, Jing and {Gao}, Jianbo and {Wang}, Xingsong},
        title = "{Multifractal analysis of sunspot time series: the effects of the 11-year cycle and Fourier truncation}",
      journal = {Journal of Statistical Mechanics: Theory and Experiment},
         year = 2009,
        month = feb,
       volume = {2009},
       number = {2},
        pages = {02066},
          doi = {10.1088/1742-5468/2009/02/P02066},
       adsurl = {https://ui.adsabs.harvard.edu/abs/2009JSMTE..02..066H}
}

@ARTICLE{2011PhRvE..84b1103M,
       author = {{Movahed}, M. Sadegh and {Ghasemi}, F. and {Rahvar}, Sohrab and {Tabar}, M. Reza Rahimi},
        title = "{Long-range correlation in cosmic microwave background radiation}",
      journal = {\pre},
         year = 2011,
        month = aug,
       volume = {84},
       number = {2},
          eid = {021103},
        pages = {021103},
          doi = {10.1103/PhysRevE.84.021103},
archivePrefix = {arXiv},
       eprint = {astro-ph/0602461},
 primaryClass = {astro-ph},
       adsurl = {https://ui.adsabs.harvard.edu/abs/2011PhRvE..84b1103M}
}

@ARTICLE{2013MNRAS.434.3597M,
       author = {{Movahed}, M. Sadegh and {Javanmardi}, B. and {Sheth}, Ravi K.},
        title = "{Peak-peak correlations in the cosmic background radiation from cosmic strings}",
      journal = {\mnras},
         year = 2013,
        month = oct,
       volume = {434},
       number = {4},
        pages = {3597-3605},
          doi = {10.1093/mnras/stt1284},
archivePrefix = {arXiv},
       eprint = {1212.0964},
 primaryClass = {astro-ph.CO},
       adsurl = {https://ui.adsabs.harvard.edu/abs/2013MNRAS.434.3597M}
}

@ARTICLE{2014OptL...39.3718Z,
       author = {{Zunino}, Luciano and {Gulich}, Dami{\'a}n and {Funes}, Gustavo and {Ziad}, Aziz},
        title = "{Experimental confirmation of long-memory correlations in star-wander data}",
      journal = {Optics Letters},
         year = 2014,
        month = jul,
       volume = {39},
       number = {13},
        pages = {3718},
          doi = {10.1364/OL.39.003718},
archivePrefix = {arXiv},
       eprint = {1406.7706},
 primaryClass = {physics.ao-ph},
       adsurl = {https://ui.adsabs.harvard.edu/abs/2014OptL...39.3718Z}
}

@ARTICLE{2007JSMTE..04...12J,
       author = {{Jafari}, G.~R. and {Pedram}, P. and {Hedayatifar}, L.},
        title = "{Long-range correlation and multifractality in Bach's Inventions pitches}",
      journal = {Journal of Statistical Mechanics: Theory and Experiment},
         year = 2007,
        month = apr,
       volume = {2007},
       number = {4},
        pages = {04012},
          doi = {10.1088/1742-5468/2007/04/P04012},
archivePrefix = {arXiv},
       eprint = {0704.0726},
 primaryClass = {physics.data-an},
       adsurl = {https://ui.adsabs.harvard.edu/abs/2007JSMTE..04...12J}
}

@ARTICLE{1999Natur.399..461I,
       author = {{Ivanov}, Plamen Ch. and {Amaral}, Lu{\'\i}s A. Nunes and {Goldberger}, Ary L. and {Havlin}, Shlomo and {Rosenblum}, Michael G. and {Struzik}, Zbigniew R. and {Stanley}, H. Eugene},
        title = "{Multifractality in human heartbeat dynamics}",
      journal = {\nat},
         year = 1999,
        month = jun,
       volume = {399},
       number = {6735},
        pages = {461-465},
          doi = {10.1038/20924},
archivePrefix = {arXiv},
       eprint = {cond-mat/9905329},
 primaryClass = {cond-mat.stat-mech},
       adsurl = {https://ui.adsabs.harvard.edu/abs/1999Natur.399..461I}
}

@ARTICLE{2020Chaos..30g3138P,
       author = {{Pavlov}, A.~N. and {Dubrovsky}, A.~I. and {Koronovskii}, A.~A. and {Pavlova}, O.~N. and {Semyachkina-Glushkovskaya}, O.~V. and {Kurths}, J.},
        title = "{Extended detrended fluctuation analysis of electroencephalograms signals during sleep and the opening of the blood-brain barrier}",
      journal = {Chaos},
         year = 2020,
        month = jul,
       volume = {30},
       number = {7},
          eid = {073138},
        pages = {073138},
          doi = {10.1063/5.0011823},
       adsurl = {https://ui.adsabs.harvard.edu/abs/2020Chaos..30g3138P}
}

@ARTICLE{2021EPJP..136...10P,
       author = {{Pavlov}, A.~N. and {Pavlova}, O.~N. and {Semyachkina-Glushkovskaya}, O.~V. and {Kurths}, J.},
        title = "{Extended detrended fluctuation analysis: effects of nonstationarity and application to sleep data}",
      journal = {European Physical Journal Plus},
         year = 2021,
        month = jan,
       volume = {136},
       number = {1},
          eid = {10},
        pages = {10},
          doi = {10.1140/epjp/s13360-020-00980-x},
       adsurl = {https://ui.adsabs.harvard.edu/abs/2021EPJP..136...10P}
}

@ARTICLE{2020CNSNS..8505232P,
       author = {{Pavlov}, A.~N. and {Abdurashitov}, A.~S. and {Koronovskii}, A.~A. and {Pavlova}, O.~N. and {Semyachkina-Glushkovskaya}, O.~V. and {Kurths}, J.},
        title = "{Detrended fluctuation analysis of cerebrovascular responses to abrupt changes in peripheral arterial pressure in rats}",
      journal = {Communications in Nonlinear Science and Numerical Simulations},
         year = 2020,
        month = jun,
       volume = {85},
          eid = {105232},
        pages = {105232},
          doi = {10.1016/j.cnsns.2020.105232},
       adsurl = {https://ui.adsabs.harvard.edu/abs/2020CNSNS..8505232P}
}

@ARTICLE{2020SoPh..295..123L,
       author = {{Lee}, Eo-Jin and {Park}, Sung-Hong and {Moon}, Yong-Jae},
        title = "{Time Series Analysis of Photospheric Magnetic Parameters of Flare-Quiet Versus Flaring Active Regions: Scaling Properties of Fluctuations}",
      journal = {\solphys},
         year = 2020,
        month = sep,
       volume = {295},
       number = {9},
          eid = {123},
        pages = {123},
          doi = {10.1007/s11207-020-01690-4},
archivePrefix = {arXiv},
       eprint = {2008.13085},
 primaryClass = {astro-ph.SR},
       adsurl = {https://ui.adsabs.harvard.edu/abs/2020SoPh..295..123L}
}

@ARTICLE{2018ApJ...864..162E,
       author = {{Eghdami}, I. and {Panahi}, H. and {Movahed}, S.~M.~S.},
        title = "{Multifractal Analysis of Pulsar Timing Residuals: Assessment of Gravitational Wave Detection}",
      journal = {\apj},
         year = 2018,
        month = sep,
       volume = {864},
       number = {2},
          eid = {162},
        pages = {162},
          doi = {10.3847/1538-4357/aad7b9},
archivePrefix = {arXiv},
       eprint = {1704.08599},
 primaryClass = {astro-ph.SR},
       adsurl = {https://ui.adsabs.harvard.edu/abs/2018ApJ...864..162E}
}

@ARTICLE{1993ApJ...413L.101K,
       author = {{Kouveliotou}, Chryssa and {Meegan}, Charles A. and {Fishman}, Gerald J. and {Bhat}, Narayana P. and {Briggs}, Michael S. and {Koshut}, Thomas M. and {Paciesas}, William S. and {Pendleton}, Geoffrey N.},
        title = "{Identification of Two Classes of Gamma-Ray Bursts}",
      journal = {\apjl},
         year = 1993,
        month = aug,
       volume = {413},
        pages = {L101},
          doi = {10.1086/186969},
       adsurl = {https://ui.adsabs.harvard.edu/abs/1993ApJ...413L.101K}
}

@ARTICLE{2006ARA&A..44..507W,
       author = {{Woosley}, S.~E. and {Bloom}, J.~S.},
        title = "{The Supernova Gamma-Ray Burst Connection}",
      journal = {\araa},
         year = 2006,
        month = sep,
       volume = {44},
       number = {1},
        pages = {507-556},
          doi = {10.1146/annurev.astro.43.072103.150558},
archivePrefix = {arXiv},
       eprint = {astro-ph/0609142},
 primaryClass = {astro-ph},
       adsurl = {https://ui.adsabs.harvard.edu/abs/2006ARA&A..44..507W}
}

@ARTICLE{2017ApJ...848L..13A,
       author = {{Abbott}, B.~P. and {Abbott}, R. and {Abbott}, T.~D. and {Acernese}, F. and {Ackley}, K. and {Adams}, C. and {Adams}, T. and {Addesso}, P. and {Adhikari}, R.~X. and {Adya}, V.~B. and {Affeldt}, C. and {Afrough}, M. and {Agarwal}, B. and {Agathos}, M. and {Agatsuma}, K. and {Aggarwal}, N. and {Aguiar}, O.~D. and {Aiello}, L. and {Ain}, A. and {Ajith}, P. and {Allen}, B. and {Allen}, G. and {Allocca}, A. and {Aloy}, M.~A. and {Altin}, P.~A. and {Amato}, A. and {Ananyeva}, A. and {Anderson}, S.~B. and {Anderson}, W.~G. and {Angelova}, S.~V. and {Antier}, S. and {Appert}, S. and {Arai}, K. and {Araya}, M.~C. and {Areeda}, J.~S. and {Arnaud}, N. and {Arun}, K.~G. and {Ascenzi}, S. and {Ashton}, G. and {Ast}, M. and {Aston}, S.~M. and {Astone}, P. and {Atallah}, D.~V. and {Aufmuth}, P. and {Aulbert}, C. and {AultONeal}, K. and {Austin}, C. and {Avila-Alvarez}, A. and {Babak}, S. and {Bacon}, P. and {Bader}, M.~K.~M. and {Bae}, S. and {Baker}, P.~T. and {Baldaccini}, F. and {Ballardin}, G. and {Ballmer}, S.~W. and {Banagiri}, S. and {Barayoga}, J.~C. and {Barclay}, S.~E. and {Barish}, B.~C. and {Barker}, D. and {Barkett}, K. and {Barone}, F. and {Barr}, B. and {Barsotti}, L. and {Barsuglia}, M. and {Barta}, D. and {Bartlett}, J. and {Bartos}, I. and {Bassiri}, R. and {Basti}, A. and {Batch}, J.~C. and {Bawaj}, M. and {Bayley}, J.~C. and {Bazzan}, M. and {B{\'e}csy}, B. and {Beer}, C. and {Bejger}, M. and {Belahcene}, I. and {Bell}, A.~S. and {Berger}, B.~K. and {Bergmann}, G. and {Bero}, J.~J. and {Berry}, C.~P.~L. and {Bersanetti}, D. and {Bertolini}, A. and {Betzwieser}, J. and {Bhagwat}, S. and {Bhandare}, R. and {Bilenko}, I.~A. and {Billingsley}, G. and {Billman}, C.~R. and {Birch}, J. and {Birney}, R. and {Birnholtz}, O. and {Biscans}, S. and {Biscoveanu}, S. and {Bisht}, A. and {Bitossi}, M. and {Biwer}, C. and {Bizouard}, M.~A. and {Blackburn}, J.~K. and {Blackman}, J. and {Blair}, C.~D. and {Blair}, D.~G. and {Blair}, R.~M. and {Bloemen}, S. and {Bock}, O. and {Bode}, N. and {Boer}, M. and {Bogaert}, G. and {Bohe}, A. and {Bondu}, F. and {Bonilla}, E. and {Bonnand}, R. and {Boom}, B.~A. and {Bork}, R. and {Boschi}, V. and {Bose}, S. and {Bossie}, K. and {Bouffanais}, Y. and {Bozzi}, A. and {Bradaschia}, C. and {Brady}, P.~R. and {Branchesi}, M. and {Brau}, J.~E. and {Briant}, T. and {Brillet}, A. and {Brinkmann}, M. and {Brisson}, V. and {Brockill}, P. and {Broida}, J.~E. and {Brooks}, A.~F. and {Brown}, D.~A. and {Brown}, D.~D. and {Brunett}, S. and {Buchanan}, C.~C. and {Buikema}, A. and {Bulik}, T. and {Bulten}, H.~J. and {Buonanno}, A. and {Buskulic}, D. and {Buy}, C. and {Byer}, R.~L. and {Cabero}, M. and {Cadonati}, L. and {Cagnoli}, G. and {Cahillane}, C. and {Calder{\'o}n Bustillo}, J. and {Callister}, T.~A. and {Calloni}, E. and {Camp}, J.~B. and {Canepa}, M. and {Canizares}, P. and {Cannon}, K.~C. and {Cao}, H. and {Cao}, J. and {Capano}, C.~D. and {Capocasa}, E. and {Carbognani}, F. and {Caride}, S. and {Carney}, M.~F. and {Casanueva Diaz}, J. and {Casentini}, C. and {Caudill}, S. and {Cavagli{\`a}}, M. and {Cavalier}, F. and {Cavalieri}, R. and {Cella}, G. and {Cepeda}, C.~B. and {Cerd{\'a}-Dur{\'a}n}, P. and {Cerretani}, G. and {Cesarini}, E. and {Chamberlin}, S.~J. and {Chan}, M. and {Chao}, S. and {Charlton}, P. and {Chase}, E. and {Chassande-Mottin}, E. and {Chatterjee}, D. and {Chatziioannou}, K. and {Cheeseboro}, B.~D. and {Chen}, H.~Y. and {Chen}, X. and {Chen}, Y. and {Cheng}, H. -P. and {Chia}, H. and {Chincarini}, A. and {Chiummo}, A. and {Chmiel}, T. and {Cho}, H.~S. and {Cho}, M. and {Chow}, J.~H. and {Christensen}, N. and {Chu}, Q. and {Chua}, A.~J.~K. and {Chua}, S. and {Chung}, A.~K.~W. and {Chung}, S. and {Ciani}, G.},
        title = "{Gravitational Waves and Gamma-Rays from a Binary Neutron Star Merger: GW170817 and GRB 170817A}",
      journal = {\apjl},
         year = 2017,
        month = oct,
       volume = {848},
       number = {2},
          eid = {L13},
        pages = {L13},
          doi = {10.3847/2041-8213/aa920c},
archivePrefix = {arXiv},
       eprint = {1710.05834},
 primaryClass = {astro-ph.HE},
       adsurl = {https://ui.adsabs.harvard.edu/abs/2017ApJ...848L..13A}
}

@ARTICLE{2017ApJ...848L..14G,
       author = {{Goldstein}, A. and {Veres}, P. and {Burns}, E. and {Briggs}, M.~S. and {Hamburg}, R. and {Kocevski}, D. and {Wilson-Hodge}, C.~A. and {Preece}, R.~D. and {Poolakkil}, S. and {Roberts}, O.~J. and {Hui}, C.~M. and {Connaughton}, V. and {Racusin}, J. and {von Kienlin}, A. and {Dal Canton}, T. and {Christensen}, N. and {Littenberg}, T. and {Siellez}, K. and {Blackburn}, L. and {Broida}, J. and {Bissaldi}, E. and {Cleveland}, W.~H. and {Gibby}, M.~H. and {Giles}, M.~M. and {Kippen}, R.~M. and {McBreen}, S. and {McEnery}, J. and {Meegan}, C.~A. and {Paciesas}, W.~S. and {Stanbro}, M.},
        title = "{An Ordinary Short Gamma-Ray Burst with Extraordinary Implications: Fermi-GBM Detection of GRB 170817A}",
      journal = {\apjl},
         year = 2017,
        month = oct,
       volume = {848},
       number = {2},
          eid = {L14},
        pages = {L14},
          doi = {10.3847/2041-8213/aa8f41},
archivePrefix = {arXiv},
       eprint = {1710.05446},
 primaryClass = {astro-ph.HE},
       adsurl = {https://ui.adsabs.harvard.edu/abs/2017ApJ...848L..14G}
}

@ARTICLE{1993ApJ...405..273W,
       author = {{Woosley}, S.~E.},
        title = "{Gamma-Ray Bursts from Stellar Mass Accretion Disks around Black Holes}",
      journal = {\apj},
         year = 1993,
        month = mar,
       volume = {405},
        pages = {273},
          doi = {10.1086/172359},
       adsurl = {https://ui.adsabs.harvard.edu/abs/1993ApJ...405..273W}
}

@ARTICLE{1998ApJ...494L..45P,
       author = {{Paczy{\'n}ski}, Bohdan},
        title = "{Are Gamma-Ray Bursts in Star-Forming Regions?}",
      journal = {\apjl},
         year = 1998,
        month = feb,
       volume = {494},
       number = {1},
        pages = {L45-L48},
          doi = {10.1086/311148},
archivePrefix = {arXiv},
       eprint = {astro-ph/9710086},
 primaryClass = {astro-ph},
       adsurl = {https://ui.adsabs.harvard.edu/abs/1998ApJ...494L..45P}
}

@ARTICLE{1992ApJ...397..570M,
       author = {{Meszaros}, P. and {Rees}, M.~J.},
        title = "{Tidal Heating and Mass Loss in Neutron Star Binaries: Implications for Gamma-Ray Burst Models}",
      journal = {\apj},
         year = 1992,
        month = oct,
       volume = {397},
        pages = {570},
          doi = {10.1086/171813},
       adsurl = {https://ui.adsabs.harvard.edu/abs/1992ApJ...397..570M}
}

@ARTICLE{1998ApJ...507L..59L,
       author = {{Li}, Li-Xin and {Paczy{\'n}ski}, Bohdan},
        title = "{Transient Events from Neutron Star Mergers}",
      journal = {\apjl},
         year = 1998,
        month = nov,
       volume = {507},
       number = {1},
        pages = {L59-L62},
          doi = {10.1086/311680},
archivePrefix = {arXiv},
       eprint = {astro-ph/9807272},
 primaryClass = {astro-ph},
       adsurl = {https://ui.adsabs.harvard.edu/abs/1998ApJ...507L..59L}
}

@ARTICLE{1986ApJ...308L..43P,
       author = {{Paczynski}, B.},
        title = "{Gamma-ray bursters at cosmological distances}",
      journal = {\apjl},
         year = 1986,
        month = sep,
       volume = {308},
        pages = {L43-L46},
          doi = {10.1086/184740},
       adsurl = {https://ui.adsabs.harvard.edu/abs/1986ApJ...308L..43P}
}

@ARTICLE{1991AcA....41..257P,
       author = {{Paczynski}, Bohdan},
        title = "{Cosmological gamma-ray bursts.}",
      journal = {\actaa},
         year = 1991,
        month = jan,
       volume = {41},
        pages = {257-267},
       adsurl = {https://ui.adsabs.harvard.edu/abs/1991AcA....41..257P}
}

@ARTICLE{2011ApJ...726...90Z,
       author = {{Zhang}, Bing and {Yan}, Huirong},
        title = "{The Internal-collision-induced Magnetic Reconnection and Turbulence (ICMART) Model of Gamma-ray Bursts}",
      journal = {\apj},
         year = 2011,
        month = jan,
       volume = {726},
       number = {2},
          eid = {90},
        pages = {90},
          doi = {10.1088/0004-637X/726/2/90},
archivePrefix = {arXiv},
       eprint = {1011.1197},
 primaryClass = {astro-ph.HE},
       adsurl = {https://ui.adsabs.harvard.edu/abs/2011ApJ...726...90Z}
}

@ARTICLE{2012MNRAS.425L..32M,
       author = {{MacLachlan}, G.~A. and {Shenoy}, A. and {Sonbas}, E. and {Dhuga}, K.~S. and {Eskandarian}, A. and {Maximon}, L.~C. and {Parke}, W.~C.},
        title = "{The minimum variability time-scale and its relation to pulse profiles of Fermi GRBs}",
      journal = {\mnras},
         year = 2012,
        month = sep,
       volume = {425},
       number = {1},
        pages = {L32-L35},
          doi = {10.1111/j.1745-3933.2012.01295.x},
archivePrefix = {arXiv},
       eprint = {1205.0055},
 primaryClass = {astro-ph.HE},
       adsurl = {https://ui.adsabs.harvard.edu/abs/2012MNRAS.425L..32M}
}

@ARTICLE{1993ApJ...413..281B,
       author = {{Band}, D. and {Matteson}, J. and {Ford}, L. and {Schaefer}, B. and {Palmer}, D. and {Teegarden}, B. and {Cline}, T. and {Briggs}, M. and {Paciesas}, W. and {Pendleton}, G. and {Fishman}, G. and {Kouveliotou}, C. and {Meegan}, C. and {Wilson}, R. and {Lestrade}, P.},
        title = "{BATSE Observations of Gamma-Ray Burst Spectra. I. Spectral Diversity}",
      journal = {\apj},
         year = 1993,
        month = aug,
       volume = {413},
        pages = {281},
          doi = {10.1086/172995},
       adsurl = {https://ui.adsabs.harvard.edu/abs/1993ApJ...413..281B}
}

@ARTICLE{2021ApJ...919...37H,
       author = {{Hakkila}, Jon},
        title = "{How Temporal Symmetry Defines Morphology in BATSE Gamma-Ray Burst Pulse Light Curves}",
      journal = {\apj},
         year = 2021,
        month = sep,
       volume = {919},
       number = {1},
          eid = {37},
        pages = {37},
          doi = {10.3847/1538-4357/ac110c},
archivePrefix = {arXiv},
       eprint = {2108.13937},
 primaryClass = {astro-ph.HE},
       adsurl = {https://ui.adsabs.harvard.edu/abs/2021ApJ...919...37H}
}

@ARTICLE{Fenimore2000,
   author = {{Fenimore}, E.~E. and {Ramirez-Ruiz}, E.},
    title = "{Redshifts For 220 BATSE Gamma‑Ray Bursts Determined by Variability and the Cosmological Consequences}",
  journal = {arXiv e-prints},
     year = 2000,
      eprint = {astro-ph/0004176},
}

@ARTICLE{Beloborodov2000,
   author = {{Beloborodov}, A.~M. and {Stern}, B.~E. and {Svensson}, R.},
    title = "{Power Density Spectra of Gamma‑Ray Bursts}",
  journal = {\apj},
     year = 2000,
    volume = 535,
    pages = {158-164},
      doi = {10.1086/308834},
}

@ARTICLE{Borgonovo2004,
       author = {{Borgonovo}, L.},
        title = "{Bimodal distribution of the autocorrelation function in gamma-ray bursts}",
      journal = {\aap},
         year = 2004,
        month = may,
       volume = {418},
        pages = {487-493},
          doi = {10.1051/0004-6361:20034567},
archivePrefix = {arXiv},
       eprint = {astro-ph/0402107},
 primaryClass = {astro-ph},
       adsurl = {https://ui.adsabs.harvard.edu/abs/2004A&A...418..487B}
}

@ARTICLE{Reichart2001,
   author = {{Reichart}, D.~E. and {Lamb}, D.~Q. and {Fenimore}, E.~E. and {Ramirez-Ruiz}, E. and 
             {Cline}, T.~L. and {Hurley}, K. and {Czerny}, B.},
    title = "{A Possible Cepheid‑like Luminosity Estimator for the Long Gamma‑Ray Bursts}",
  journal = {\apj},
     year = 2001,
    volume = 552,
     pages = {57-71},
      doi = {10.1086/320255},
}

@ARTICLE{Guidorzi2005,
       author = {{Guidorzi}, C. and {Frontera}, F. and {Montanari}, E. and {Rossi}, F. and {Amati}, L. and {Gomboc}, A. and {Hurley}, K. and {Mundell}, C.~G.},
        title = "{The gamma-ray burst variability-peak luminosity correlation: new results}",
      journal = {\mnras},
         year = 2005,
        month = oct,
       volume = {363},
       number = {1},
        pages = {315-325},
          doi = {10.1111/j.1365-2966.2005.09450.x},
archivePrefix = {arXiv},
       eprint = {astro-ph/0507588},
 primaryClass = {astro-ph},
       adsurl = {https://ui.adsabs.harvard.edu/abs/2005MNRAS.363..315G}
}

@ARTICLE{Norris1996,
       author = {{Norris}, J.~P. and {Nemiroff}, R.~J. and {Bonnell}, J.~T. and {Scargle}, J.~D. and {Kouveliotou}, C. and {Paciesas}, W.~S. and {Meegan}, C.~A. and {Fishman}, G.~J.},
        title = "{Attributes of Pulses in Long Bright Gamma-Ray Bursts}",
      journal = {\apj},
         year = 1996,
        month = mar,
       volume = {459},
        pages = {393},
          doi = {10.1086/176902},
       adsurl = {https://ui.adsabs.harvard.edu/abs/1996ApJ...459..393N}
}

@ARTICLE{Wang2020,
       author = {{Wang}, Feifei and {Zou}, Yuan-Chuan and {Liu}, Fuxiang and {Liao}, Bin and {Liu}, Yu and {Chai}, Yating and {Xia}, Lei},
        title = "{A Comprehensive Statistical Study of Gamma-Ray Bursts}",
      journal = {\apj},
         year = 2020,
        month = apr,
       volume = {893},
       number = {1},
          eid = {77},
        pages = {77},
          doi = {10.3847/1538-4357/ab0a86},
archivePrefix = {arXiv},
       eprint = {1902.05489},
 primaryClass = {astro-ph.HE},
       adsurl = {https://ui.adsabs.harvard.edu/abs/2020ApJ...893...77W}
}

@ARTICLE{2000astro.ph..4176F,
       author = {{Fenimore}, E.~E. and {Ramirez-Ruiz}, E.},
        title = "{Redshifts For 220 BATSE Gamma-Ray Bursts Determined by Variability and the Cosmological Consequences}",
      journal = {arXiv e-prints},
         year = 2000,
        month = apr,
          eid = {astro-ph/0004176},
        pages = {astro-ph/0004176},
          doi = {10.48550/arXiv.astro-ph/0004176},
archivePrefix = {arXiv},
       eprint = {astro-ph/0004176},
 primaryClass = {astro-ph},
       adsurl = {https://ui.adsabs.harvard.edu/abs/2000astro.ph..4176F}
}

@ARTICLE{2022CoPhC.27308254R,
       author = {{Rydin Gorj{\~a}o}, Leonardo and {Hassan}, Galib and {Kurths}, J{\"u}rgen and {Witthaut}, Dirk},
        title = "{MFDFA: Efficient multifractal detrended fluctuation analysis in python}",
      journal = {Computer Physics Communications},
         year = 2022,
        month = apr,
       volume = {273},
          eid = {108254},
        pages = {108254},
          doi = {10.1016/j.cpc.2021.108254},
archivePrefix = {arXiv},
       eprint = {2104.10470},
 primaryClass = {physics.comp-ph},
       adsurl = {https://ui.adsabs.harvard.edu/abs/2022CoPhC.27308254R}
}

@ARTICLE{1895RSPS...58..240P,
       author = {{Pearson}, Karl},
        title = "{Note on Regression and Inheritance in the Case of Two Parents}",
      journal = {Proceedings of the Royal Society of London Series I},
         year = 1895,
        month = jan,
       volume = {58},
        pages = {240-242},
       adsurl = {https://ui.adsabs.harvard.edu/abs/1895RSPS...58..240P}
}

@article{Hurst1951,
  author = {{Hurst}, Harold E.},
  title={Long-term storage capacity of reservoirs},
  journal={Transactions of the American society of civil engineers},
  volume={116},
  number={1},
  pages={770--799},
  year={1951},
  publisher={American Society of Civil Engineers}
}

@ARTICLE{2023ApJ...949L..33W,
       author = {{Wang}, F.~Y. and {Wu}, Q. and {Dai}, Z.~G.},
        title = "{Repeating Fast Radio Bursts Reveal Memory from Minutes to an Hour}",
      journal = {\apjl},
         year = 2023,
        month = jun,
       volume = {949},
       number = {2},
          eid = {L33},
        pages = {L33},
          doi = {10.3847/2041-8213/acd5d2},
archivePrefix = {arXiv},
       eprint = {2302.06802},
 primaryClass = {astro-ph.HE},
       adsurl = {https://ui.adsabs.harvard.edu/abs/2023ApJ...949L..33W}
}

@ARTICLE{1995ApJ...439..542N,
       author = {{Norris}, J.~P. and {Bonnell}, J.~T. and {Nemiroff}, R.~J. and {Scargle}, J.~D. and {Kouveliotou}, C. and {Paciesas}, W.~S. and {Meegan}, C.~A. and {Fishman}, G.~J.},
        title = "{Duration Distributions of Bright and DIM BATSE Gamma-Ray Bursts}",
      journal = {\apj},
         year = 1995,
        month = feb,
       volume = {439},
        pages = {542},
          doi = {10.1086/175194},
archivePrefix = {arXiv},
       eprint = {astro-ph/9408063},
 primaryClass = {astro-ph},
       adsurl = {https://ui.adsabs.harvard.edu/abs/1995ApJ...439..542N}
}

@ARTICLE{2020ApJS..250....1T,
       author = {{Tarnopolski}, Mariusz and {{\.Z}ywucka}, Natalia and {Marchenko}, Volodymyr and {Pascual-Granado}, Javier},
        title = "{A Comprehensive Power Spectral Density Analysis of Astronomical Time Series. I. The Fermi-LAT Gamma-Ray Light Curves of Selected Blazars}",
      journal = {\apjs},
         year = 2020,
        month = sep,
       volume = {250},
       number = {1},
          eid = {1},
        pages = {1},
          doi = {10.3847/1538-4365/aba2c7},
archivePrefix = {arXiv},
       eprint = {2006.03991},
 primaryClass = {astro-ph.HE},
       adsurl = {https://ui.adsabs.harvard.edu/abs/2020ApJS..250....1T}
}

@ARTICLE{2021ApJ...911...20T,
       author = {{Tarnopolski}, Mariusz and {Marchenko}, Volodymyr},
        title = "{A Comprehensive Power Spectral Density Analysis of Astronomical Time Series. II. The Swift/BAT Long Gamma-Ray Bursts}",
      journal = {\apj},
         year = 2021,
        month = apr,
       volume = {911},
       number = {1},
          eid = {20},
        pages = {20},
          doi = {10.3847/1538-4357/abe5b1},
archivePrefix = {arXiv},
       eprint = {2102.05330},
 primaryClass = {astro-ph.HE},
       adsurl = {https://ui.adsabs.harvard.edu/abs/2021ApJ...911...20T}
}

@ARTICLE{2002A&A...390...81A,
       author = {{Amati}, L. and {Frontera}, F. and {Tavani}, M. and {in't Zand}, J.~J.~M. and {Antonelli}, A. and {Costa}, E. and {Feroci}, M. and {Guidorzi}, C. and {Heise}, J. and {Masetti}, N. and {Montanari}, E. and {Nicastro}, L. and {Palazzi}, E. and {Pian}, E. and {Piro}, L. and {Soffitta}, P.},
        title = "{Intrinsic spectra and energetics of BeppoSAX Gamma-Ray Bursts with known redshifts}",
      journal = {\aap},
         year = 2002,
        month = jul,
       volume = {390},
        pages = {81-89},
          doi = {10.1051/0004-6361:20020722},
archivePrefix = {arXiv},
       eprint = {astro-ph/0205230},
 primaryClass = {astro-ph},
       adsurl = {https://ui.adsabs.harvard.edu/abs/2002A&A...390...81A}
}

@ARTICLE{2004ApJ...616..331G,
       author = {{Ghirlanda}, Giancarlo and {Ghisellini}, Gabriele and {Lazzati}, Davide},
        title = "{The Collimation-corrected Gamma-Ray Burst Energies Correlate with the Peak Energy of Their {\ensuremath{\nu}}F$_{{\ensuremath{\nu}}}$ Spectrum}",
      journal = {\apj},
         year = 2004,
        month = nov,
       volume = {616},
       number = {1},
        pages = {331-338},
          doi = {10.1086/424913},
archivePrefix = {arXiv},
       eprint = {astro-ph/0405602},
 primaryClass = {astro-ph},
       adsurl = {https://ui.adsabs.harvard.edu/abs/2004ApJ...616..331G}
}

\end{document}